\documentclass[twocolumn,English,aps,pra,10pt,superscriptaddress,floatfix]{revtex4-2}
\usepackage{times}
\usepackage{graphicx}
\usepackage{amssymb}
\usepackage{bm}% bold math
\usepackage{amssymb,amsfonts,amsmath,amsbsy,bm,t1enc,latexsym}
\usepackage{float}
\usepackage[colorlinks=true,citecolor=blue,linkcolor=magenta]{hyperref}
\usepackage[markup=blue, authormarkupposition=left]{changes}
\usepackage{url}
\usepackage{xspace} 
\usepackage{siunitx}
\DeclareSIUnit \dbc {dBc}
\usepackage{soul}
\usepackage{changes}
\usepackage{cancel}
\usepackage{times}
\usepackage{soul}
\usepackage{makecell}
 
\usepackage{amsmath}
\usepackage{hyperref}

\usepackage{booktabs}

\newcommand{\ignore}[1]{}

\newcommand{\SiN}[0]{$\mathrm{Si_{3}N_{4}}$}

\begin{document}

\title{Dual-Bus Resonator for Multi-Port Spectral Engineering}

\author{Taewon Kim}\thanks{These authors contributed equally.}
\affiliation{School of Electrical Engineering, Korea Advanced Institute of Science and Technology, Daejeon 34141, Republic of Korea}

\author{Mehedi Hasan}\thanks{These authors contributed equally.}
\affiliation{Department of Electrical and Computer Engineering, Texas Tech University, Lubbock, TX 79409, USA}

\author{Yu Sung Choi}
\affiliation{Department of Physics, Hanyang University, Seoul 04763, Republic of Korea}

\author{Jae Woong Yoon}
\email[]{yoonjw@hanyang.ac.kr}
\affiliation{Department of Physics, Hanyang University, Seoul 04763, Republic of Korea}

\author{Sangsik Kim}
\email[]{sangsik.kim@kaist.ac.kr}
\affiliation{School of Electrical Engineering, Korea Advanced Institute of Science and Technology, Daejeon 34141, Republic of Korea}
% \affiliation{Graduate School of Quantum Science and Technology, Korea Advanced Institute of Science and Technology, Daejeon 34141, Republic of Korea}

\begin{abstract}
Microresonators are essential in integrated photonics, enabling optical filters, modulators, sensors, and frequency converters. 
Their spectral response is governed by bus-to-resonator coupling, typically classified as under-, critical-, or over-coupling.
Conventional single-bus designs inevitably link the conditions for critical coupling, a transmission zero, and maximum intra-cavity power, preventing independent control of these phenomena and restricting the ability to engineer coupling regimes and resonance lineshapes.
Here we propose and experimentally demonstrate a dual-bus racetrack resonator that breaks this constraint.
Our design demonstrates complementary channel-specific coupling regimes and enables wavelength-dependent Lorentzian-to-Fano lineshaping.
We model the device using three-waveguide coupled-mode theory and pole-zero analysis, which reveals that transmission zeros are decoupled from cavity-defined critical coupling and maximum intra-cavity power.
Furthermore, the dual-bus scheme operates broadband, spanning visible to mid-infrared across all four transmission channels, highlighting its spectral richness and platform independence.  
These results establish a general framework for multi-port spectral engineering in integrated photonics, with broad implications for tunable filters, modulators, sensors, and nonlinear optical systems.
\end{abstract}
\maketitle
%%%%%%%%%%%%%%%%%%%%%%%%%%%%%%%%%%%%%%%%%%%%%%%%%%%%%%%%%%%%%%%%%%%%%%%%%%%%%%%%%%%%%%%%%%%%%%%%%%%%%%%%%%%%%%%%%%%%%%

\section{Introduction}
Microresonators are fundamental building blocks of integrated photonics and have played a central role.  
With tunable spectral filtering and high-speed response, microresonator-based optical filters~\cite{Madsen1998,Dong:10,Wang:22} and modulators~\cite{Xu2005,Hu2023,Geravand2025,Xu:06,YuanYuan2024,Sacher:13} have become key enablers of on-chip optical communications \cite{Bogaerts2012,Hsu2023,Pirmoradi:25,Daudlin:25}.  
Their small modal volume and low optical loss also enhance light--matter interactions, making them ideal platforms for frequency comb generation~\cite{KippenbergDKS2018,Xuan:16,Stern2018}, on-chip lasing~\cite{Li2022,Ko2025,Kazakov2024,Zhao:18}, and various nonlinear~\cite{Spencer2018,Sayson2019} and quantum applications~\cite{Wu:21,Li:25,Jahanbozorgi:23,Shen:25}.
Additionally, extreme sensitivity to the local refractive index has established them as a leading platform for high-performance chemical and biological sensing~\cite{Ksendzov:05,Liu:17,Wang:25,Guo24}.

\begin{figure}[!t]
  \centering
  \includegraphics[width=0.90\columnwidth]{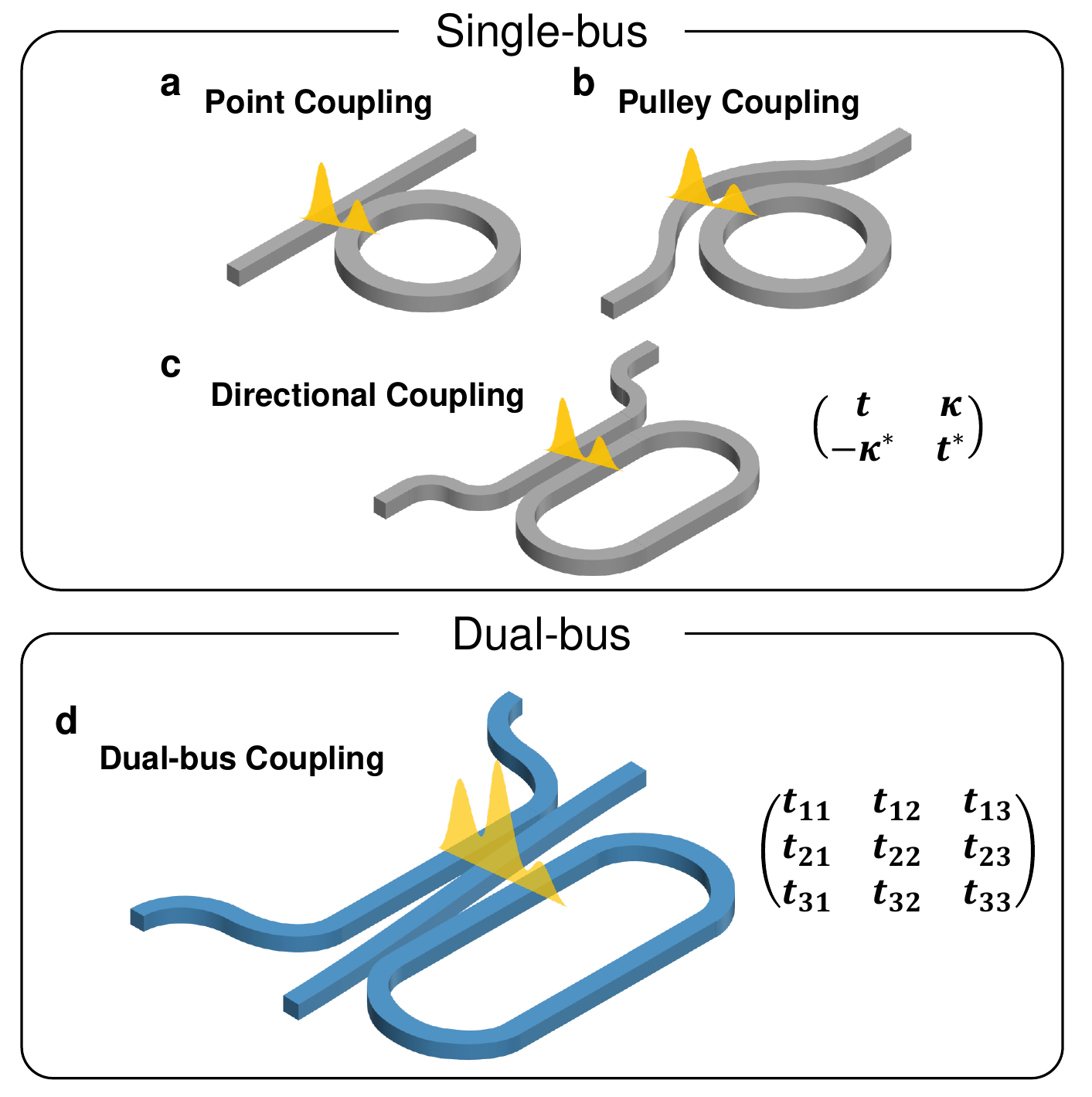}
 \caption{\textbf{Bus-to-resonator coupling configurations.} 
    \textbf{a}, Point coupling (straight coupling).
    \textbf{b}, Pulley coupling.
    \textbf{c}, Directional coupling.
    \textbf{d}, Dual-bus coupling (this work).
    While \textbf{a–c} exhibit single bus–resonator interactions, the proposed scheme in \textbf{d} introduces dual-bus interactions with the resonator.
   }
  \label{fig1}
  %\vspace{-10pt}
\end{figure}

A key factor governing microresonator response is bus-to-resonator coupling, conventionally classified as under-, over-, or critical coupling depending on the balance between coupling and intrinsic losses~\cite{yariv2000,Cai2000,Soltani2010}.
Considerable research has focused on engineering this interaction through various schemes, such as point, pulley, and directional coupling. 
Point coupling (straight coupling, Fig.~\ref{fig1}\textbf{a}) is the most classical scheme, where the interaction is very short at a point.  
It is structurally simple, but is highly sensitive to the bus-resonator gap, making it vulnerable to fabrication errors and limiting bandwidth \cite{Bogaerts2012,Tseng:13,sun2025,Li2022}.
Pulley coupling (Fig.~\ref{fig1}\textbf{b}), where the bus waveguide partially wraps around the resonator, can increase coupling bandwidths by leveraging the phase mismatch between bus and resonator modes \cite{Moille:19,Hosseini:10}.
This scheme has been widely used for small-radius rings in broadband microcombs \cite{Song2024,Moille2021,Kim2017,Moille:18,Li2016} and high-$Q$ mode coupling \cite{Spencer:14,Yang2018}.
Directional coupling (Fig.~\ref{fig1}\textbf{c}) is essentially similar to a pulley, except that it is realized with parallel waveguides.
Two parallel waveguides are typically phase-matched and exhibit a relatively strong spectral variation of transmission coefficient $t(\lambda)$, leading to periodic transitions between under- and over-coupled regimes.
It has been applied in resonator-based lasers, modulators, and driven-soliton generation \cite{Li2022,YuanYuan2024,Kazakov2025}.
Other methods, such as two-point coupling, have also been explored for tuning coupling regimes over a broad bandwidth \cite{Liu:24,Hwang2025}.

Despite their diversity, all these schemes fundamentally rely on a single bus-resonator interaction. 
This reliance leads to a consequence often considered obvious but rarely challenged: several key conditions---critical coupling (balance of intrinsic and coupling losses), a transmission zero (spectral interference null), and maximum intra-cavity power (peak energy enhancement)---all coincide at the same operating point. 
While this coincidence provides a well-defined design target, it prevents independent control of these phenomena and limits flexibility in tailoring $t(\lambda)$, coupling regimes, and resonance lineshapes.

In this paper, we introduce a dual-bus racetrack resonator with an asymmetric three-waveguide coupling, where two bus waveguides simultaneously interact with a single resonator on the same side (Fig.~\ref{fig1}\textbf{d}).
This configuration expands beyond conventional single-bus coupling, enabling complementary channel-defined coupling regimes, Lorentzian-Fano lineshape transitions, and decoupling of critical coupling, transmission zero, and intra-cavity power conditions.
We model this configuration using three-waveguide coupled-mode equations and pole-zero analysis, supported by experiments, which together establish coupling regimes that accurately predict complex spectral responses.
Moreover, multi-port interference gives rise to broadband Lorentzian-Fano transitions, with resonances evolving from symmetric Lorentzian to strongly asymmetric Fano lineshapes.
More importantly, we show that critical coupling, transmission zero, and maximum intra-cavity power---traditionally identical in single-bus resonators---are separated into distinct phenomena occurring at different wavelengths.

%%%%%%%%%%%%%%%%%%%%%%%%%%%%%%%%%%%%%%%%
%%%%%%%%%%%%%%%%%%%%%%%%%%%%%%%%%%%%%%%%

\section{Theory}
%%%%%%%%%%%%%%%%%%%%%%%%%%%%%%%
\begin{figure*}[!th]
	\centering
	\includegraphics[width=0.90\linewidth]{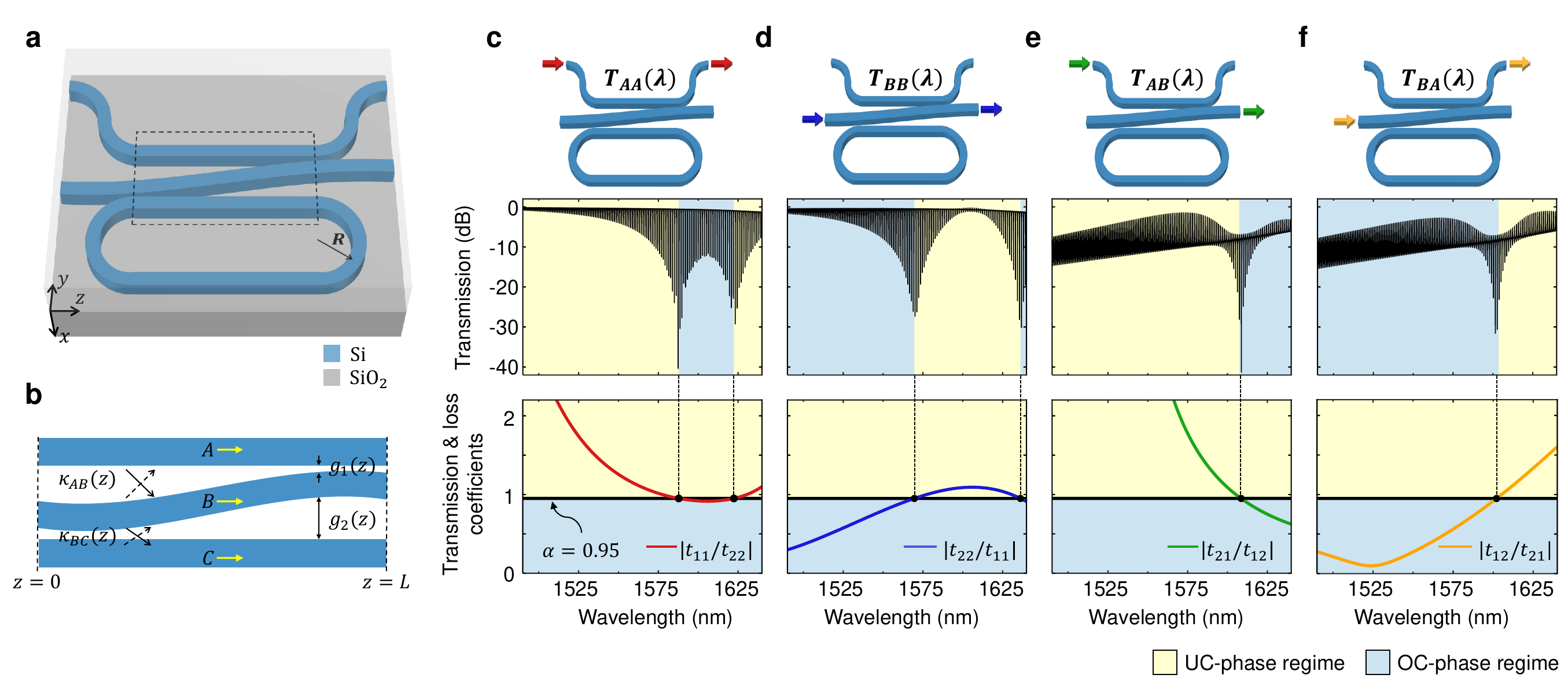}
	\caption{\textbf{ 
    Schematic of the dual-bus racetrack resonator and its transmission spectra.}
        \textbf{a}, Schematic of the proposed device.  
        \textbf{b}, Top view of the coupling region, consisting of a three-waveguide directional coupler; waveguide $B$ is bent, varying gaps $g_1$ and $g_2$, and thus coupling coefficients $\kappa_{AB}$ and $\kappa_{BC}$ along $z$.
        \textbf{c-f}, Simulated transmission spectra for four input-output configurations: 
        \textbf{c}, $T_{AA}$,
        \textbf{d}, $T_{BB}$,
        \textbf{e}, $T_{AB}$, and
        \textbf{f}, $T_{BA}$.
    The loss coefficient is set to $\alpha=0.95$ (black lines). 
    Coupling regimes are determined by the relative magnitude between $\alpha$ and the transmission coefficient ratios $t_{11}/t_{22}$ ($T_{AA}$, red), $t_{22}/t_{11}$ ($T_{BB}$, blue), $t_{21}/t_{12}$ ($T_{AB}$, green), and $t_{12}/t_{21}$ ($T_{BA}$, yellow).
    The yellow and blue shaded regions represent channel-defined under- and over-coupling-phase regimes, respectively.
}
	\label{fig2}
\end{figure*}
%%%%%%%%%%%%%%%%%%%%%%%%%%%%%%%

Figure~\ref{fig2}$\textbf{a}$ illustrates the proposed dual-bus racetrack resonator, 
which consists of two bus waveguides: a straight upper waveguide $A$ and a bent lower waveguide $B$. 
Waveguide $B$ is also coupled to the straight section of the racetrack resonator, which is denoted as waveguide $C$. 
This structure can be modeled as a lossless three-waveguide directional coupler as shown in Fig.~\ref{fig2}$\textbf{b}$, with the resonator boundary conditions imposed on waveguide C. 
All waveguides, including the resonator, have identical cross-sections that are designed to support the fundamental transverse electric (TE) mode.
Waveguide $B$ is bent according to a hyperbolic cosine function (see Section 1 of Supplementary Information), causing the gaps $g_1(z)$ between waveguides $A$ and $B$, and $g_2(z)$ between waveguides $B$ and $C$, to vary smoothly along the coupling length $L$. 

\subsection{Coupled mode theory for dual-bus coupling}
We denote the complex modal amplitudes in each waveguide as $A(z)$, $B(z)$, and $C(z)$. 
Under the weak coupling assumption \cite{yariv2007photonics}, the coupled-mode equation for the lossless three-waveguide directional coupler can be expressed as \cite{PASPALAKIS200630}:
\begin{equation}
\frac{\partial}{\partial z}
\begin{pmatrix}
A(z) \\
B(z) \\
C(z)
\end{pmatrix}
=
i
\begin{pmatrix}
0 & \kappa_{AB} & 0 \\
\kappa_{AB} & \Delta\beta& \kappa_{BC} \\
0 & \kappa_{BC} & 0
\end{pmatrix}
\begin{pmatrix}
A(z) \\
B(z) \\
C(z)
\end{pmatrix},
\label{eq:CME}
\end{equation}
where $\kappa_{AB}(z)$ and $\kappa_{BC}(z)$ are the coupling coefficients 
between the waveguides $A$ and $B$ and between $B$ and $C$, respectively, 
and $\Delta\beta=\beta_{B}-\beta$ is a propagation constant difference between waveguides $B$ ($\beta_B$) and the other two waveguides ($\beta=\beta_A=\beta_C$) due to bending of waveguide $B$.
Solving Eq.~(\ref{eq:CME}), the transfer matrix of three-waveguide coupler (denoted as $\mathbf{T}$) is a $3 \times 3$ para-unitary matrix whose elements $t_{ij}$ represent the transmission coefficients \cite{Poon:04,popovic2008phdthesis}:
%%%%%%%%%%%%%%%%%%%%%%%%%%%
\begin{equation}
\begin{pmatrix}
A(L) \\
B(L) \\
C(L)
\end{pmatrix}
=
\begin{pmatrix}
t_{11} & t_{12} & t_{13} \\
t_{21} & t_{22} & t_{23} \\
t_{31} & t_{32} & t_{33}
\end{pmatrix}
\begin{pmatrix}
A(0) \\
B(0) \\
C(0)
\end{pmatrix}.
\label{eq:TMM}
\end{equation}
%%%%%%%%%%%%%%%%%%%%%%%%%%%
Since the dual-bus racetrack resonator has two inputs and two outputs, four transfer functions $t_{AA}$, $t_{BB}$, $t_{AB}$, and $t_{BA}$ are defined based on its input-output configurations~\cite{Le2013}: 
%%%%%%%%%%%%%%%%%%%%%%%%%%%%%%%%%%%%%%%%%%%%%%%
\begin{subequations}
\begin{align}
t_{AA}
&=
\frac{A(L)}{A(0)}
=
\frac{t_{11}- t_{22}^* e^{i\varphi} \alpha e^{i \theta}}{1-t_{33} \alpha e^{i\theta}},
\label{eq:simpleHAA} \\
t_{BB}
&=
\frac{B(L)}{B(0)}
=
\frac{t_{22}- t_{11}^*e^{i\varphi} \alpha e^{i \theta}}{1-t_{33} \alpha e^{i\theta}},
\label{eq:simpleHBB} \\
t_{AB}
&=
\frac{B(L)}{A(0)}
=
\frac{t_{21}- t_{12}^*e^{i\varphi} \alpha e^{i \theta}}{1-t_{33} \alpha e^{i\theta}},
\label{eq:simpleHAB} \\
t_{BA}
&=
\frac{A(L)}{B(0)}
=
\frac{t_{12}- t_{21}^*e^{i\varphi} \alpha e^{i \theta}}{1-t_{33} \alpha e^{i\theta}}.
\label{eq:simpleHBA}
\end{align}
\end{subequations}
%%%%%%%%%%%%%%%%%%%%%%%%%%%%%%%%%%%%%%%%%%%%%%%%%%%
where $\alpha$ denotes the round-trip loss coefficient (field amplitude fraction), $\theta$ the accumulated round-trip phase of the racetrack, and $\varphi$ the global phase of the three-waveguide coupler, which signifies a common phase shift experienced by light as it propagates through the coupling region.
The closed-loop relation $C(0)=C(L)\alpha e^{i\theta}$ and the para-unitary condition $\mathbf{T^{-1}}=\mathbf{T^{\dagger}}$ are used to derive the above expressions (see Section 2 of Supplementary Information).

The transmission spectra $T_{AA}$, $T_{BB}$, $T_{AB}$, and $T_{BA}$  of each input-output configuration correspond to the absolute square of the respective transfer functions, as shown in Figs.~\ref{fig2}\textbf{c}--\ref{fig2}\textbf{f}.
Note that, for a conventional single-bus resonator, the coupling regime and resonance depth of the transmission spectrum are governed by the balance between $|t|$ and $\alpha$~\cite{yariv2007photonics}.  
Analogously, in the proposed dual-bus racetrack resonator, the spectral behavior is dictated by the ratio of both transmission coefficients from each bus. 
As in \eqref{eq:simpleHAA}–\eqref{eq:simpleHBA}, the ratios $|t_{11}/t_{22}|$, $|t_{22}/t_{11}|$, $|t_{21}/t_{12}|$, and $|t_{12}/t_{21}|$ play a role similar to $|t|$ in the single-bus case, determining the coupling regime and shaping the envelope of the corresponding transmission spectra.
This behavior is illustrated in Figs.~\ref{fig2}$\textbf{c}$--\ref{fig2}$\textbf{f}$, where each transmission spectrum exhibits a distinct spectral envelope set by its wavelength-dependent transmission coefficients.
For example, in Fig.~\ref{fig2}\textbf{c}, the extinction ratio of $T_{AA}(\lambda)$ is determined by how closely the ratio $|t_{11}/t_{22}|$ matches the resonator loss coefficient, $\alpha$.
At the specific wavelengths where $|t_{11}/t_{22}|=\alpha$, the transfer function becomes zero, resulting in destructive interference at the output.
$T_{AA}(\lambda)$ exhibits prominent resonance dips with large extinction ratios, a behavior that is constantly observed in all four transmission spectra.

\subsection{Pole-zero analysis}
In a single-bus resonator, the condition for zero transmission ($\alpha = |t|$) coincides with the critical coupling, where the intrinsic and coupling loss rates are equal ($\gamma_i=\gamma_c$). 
However, this analogy is not valid in our dual-bus resonator, as direct port-port interaction affects the transmission spectra independently of the resonator.
Therefore, to analyze this more systematically, we adopt a linear-systems approach, representing each transfer function in the $z$-domain for pole-zero interpretation \cite{PhysRevB.87.125118}.
By sampling the field once per round-trip of the resonator and defining $z^{-1} = e^{i\theta}$, the transfer functions can be expressed in the $z$-domain as:
%%%%%%%%%%%%%%%%%%%%%%%%%%%%%%%%%%%%%%%%%%%%%%%
\begin{subequations}
\begin{align}
t_{AA}(z)
&=
t_{11} \frac{z - \alpha \dfrac{t_{22}^* e^{i\varphi}}{t_{11}}}{z - \alpha t_{33}},
\label{eq:ztransformHAA} \\
t_{BB}(z)
&=
t_{22}\frac{z - \alpha \dfrac{t_{11}^* e^{i\varphi}}{t_{22}}}{z - \alpha t_{33}},
\label{eq:ztransformHBB} \\
t_{AB}(z)
&=
t_{21}\frac{z - \alpha \dfrac{t_{12}^* e^{i\varphi}}{t_{21}}}{z - \alpha t_{33}},
\label{eq:ztransformHAB} \\
t_{BA}(z)
&=
t_{12}\frac{z - \alpha \dfrac{t_{21}^* e^{i\varphi}}{t_{12}}}{z - \alpha t_{33}}.
\label{eq:ztransformHBA}
\end{align}
\end{subequations}
%%%%%%%%%%%%%%%%%%%%%%%%%%%%%%%%%%%%%%%%%%%%%%%

Each transfer function is a first-order rational function, characterized by a single zero and a single pole in the complex $z$-plane. 
All four transfer functions share a common pole located at
$z_p = \alpha t_{33}$, which reflects the resonator’s round-trip amplitude attenuation. 
The magnitude of the pole directly relates to the loaded (total) amplitude decay rate $\gamma_L$ and the loaded quality factor $Q_L$, following \cite{Wang:22}:
\begin{equation}
\gamma_L = -\ln|\alpha t_{33}| /\tau,
\quad
Q_L = \frac{\omega_0}{2\gamma_L},
\label{eq:totallossandQ}
\end{equation}
where $\tau = \dfrac{n_g L_{rt}}{c}$ is the single round-trip time, $L_{rt}$ the round-trip length of the racetrack, and $\omega_0$ the resonance frequency of the racetrack resonator.

In contrast to the shared pole, each transfer function has a unique zero: 
$z_{AA} = \alpha {t_{22}^* e^{i\varphi}}/{t_{11}}$, 
$z_{BB} = \alpha {t_{11}^* e^{i\varphi}}/{t_{22}}$, 
$z_{AB} = \alpha {t_{12}^* e^{i\varphi}}/{t_{21}}$, and
$z_{BA} = \alpha {t_{21}^* e^{i\varphi}}/{t_{12}}$.

\begin{figure*}[!th]
	\centering
	\includegraphics[width=0.80\linewidth]{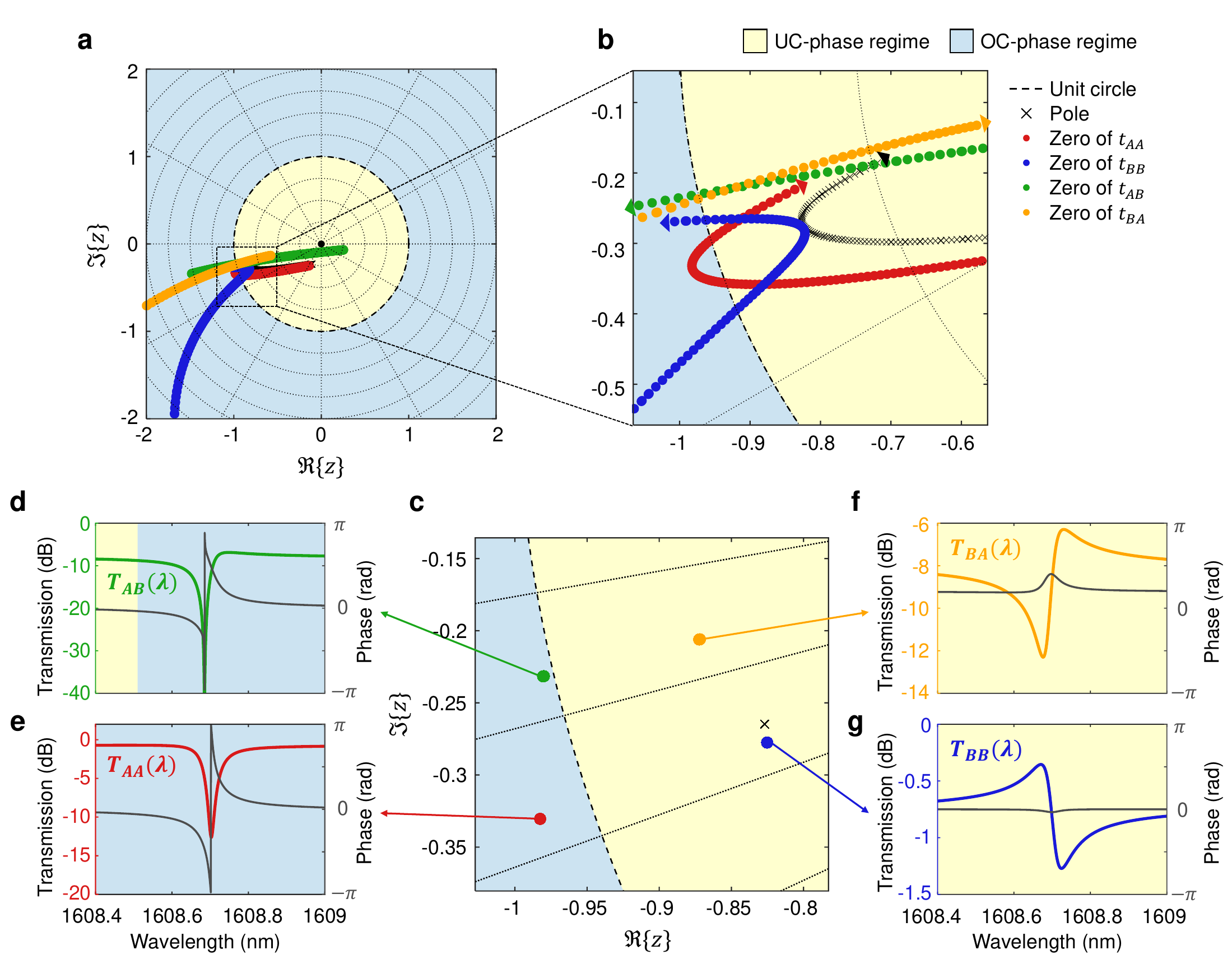}
    \caption{
	\textbf{Pole–zero analysis.}
        \textbf{a}, Pole–zero map of all four transfer functions over the wavelength range of 1490–1640~nm. 
        \textbf{b}, Zoomed-in trajectories of the pole and zeros near the unit circle; arrows indicate the direction of evolution with increasing wavelength. 
        \textbf{c}, Pole and zero locations at a resonance near 1608~nm.
        \textbf{d-g}, Representative transmission spectra and phase responses:   
        \textbf{d}, $T_{AB}(\lambda)$: Near Lorentzian dip with high extinction and sharp phase transition (transmission zero). 
        \textbf{e}, $T_{AA}(\lambda)$: Lorentzian dip with full $2\pi$ phase shift (OC phase regime). 
        \textbf{f}, $T_{BA}(\lambda)$: Fano-like lineshape due to angular separation between the pole and zero, with minimal phase variation (UC phase regime). 
        \textbf{g}, $T_{BB}(\lambda)$: Fano-like lineshape with low extinction and limited phase change.
        The Fano lineshapes in \textbf{f} and \textbf{g} are mirror-symmetric, corresponding to a sign reversal of the Fano asymmetry parameter $q$.
	}
	\label{fig3}
\end{figure*}

The corresponding wavelength-dependent pole-zero distributions for all four transfer functions are plotted in Figs.~\ref{fig3}\textbf{a} and \ref{fig3}\textbf{b}. 
The location of each zero---whether it lies on, inside, or outside the unit circle---can be determined by comparing the magnitudes of $\alpha$ to the corresponding transmission coefficient ratio, consistent with our previous analysis.

%%%%%%%%%%%%%%%%%%%%%%%%%%%%%%%%%%%%%%%%%%%%%%%
%%%%%%%%%%%%%%%%%%%%%%%%%%%%%%%%%%%%%%%%%%%%%%%

For example, Fig.~\ref{fig3}\textbf{c} presents a zoomed-in pole–zero map, where the resonances near 1608.7~nm are marked for each port, and their corresponding resonance spectra and phase responses are shown in Figs.~\ref{fig3}\textbf{d}–\ref{fig3}\textbf{g}.
When the zero lies inside the unit circle, the transfer function exhibits a phase shift less than $\pi$ across the resonance (Figs.~\ref{fig3}$\textbf{f}$, \ref{fig3}$\textbf{g}$). 
This is analogous to the phase response of an under-coupled single-bus resonator~\cite{Bogaerts2012,Twayana:21}. 
Hereinafter, we refer to it as the under-coupling-phase (UC-phase) regime.
In contrast, when the zero lies outside the unit circle, the transfer function exhibits a full $2\pi$ phase rotation (Figs.~\ref{fig3}$\textbf{d}$, \ref{fig3}$\textbf{e}$).
This corresponds to the phase response of an over-coupled single-bus resonator; thus, we refer to it as the over-coupling-phase (OC-phase) regime.
When the zero lies on the unit circle, it results in transmission zero (TZ) with an abrupt $\pi$ phase shift, similar to critical coupling in a single-bus resonator.
We use the terms UC-phase, OC-phase, and TZ to distinguish these cases from the conventional definitions of under-, over-, and critical coupling based on the resonator's power balance, associated with the loaded quality factor 
\cite{haus1984waves}. 
Here, classification is instead based on the zero's location in the pole-zero diagram, representing an interference and channel-defined perspective of the transmission response with the corresponding phase shifts.

With this classification, we can analyze the spectral behavior clearly.
The radial locations of the zeros of $t_{AA}$ and $t_{BB}$ are inter-dependent at each wavelength, constrained by $|z_{AA}||z_{BB}| = \alpha^2$. 
This implies that if the zero of $t_{AA}$ lies outside the radius-$\alpha$ circle ($|z_{AA}| > \alpha$), the zero of $t_{BB}$ must lie inside it ($|z_{BB}| < \alpha$), and vice versa. 
As the wavelength varies, $z_{AA}$ moves from inside the unit circle to outside and back, transitioning between UC-phase and OC-phase regimes, while $z_{BB}$ follows the opposite trajectory. 
This dispersive interplay gives rise to spectrally complementary responses between $T_{AA}$ and $T_{BB}$ (Figs.~\ref{fig2}$\textbf{c}$, \ref{fig2}$\textbf{d}$).
A similar complementary relationship is observed for the cross-port spectra $T_{AB}$ and $T_{BA}$, whose zeros satisfy $|z_{AB}||z_{BA}| = \alpha^2$. 
As $T_{AB}$ transitions from UC-phase to OC-phase regimes, $T_{BA}$ undergoes the opposite transition, again preserving complementarity behavior (Figs.~\ref{fig2}$\textbf{e}$, \ref{fig2}$\textbf{f})$.

\subsection{Lorentzian and Fano resonance lineshapes}
The resonance lineshapes in each port can be understood via pole-zero analysis, governed by the proximity of zero to the unit circle and the angular separation between pole and zero in the complex $z$-plane (Figs.~\ref{fig3}\textbf{d}–\ref{fig3}\textbf{g}) \cite{PhysRevB.87.125118}.
When a zero is close to the unit circle with pole and zero are nearly aligned in angle (Fig.~\ref{fig3}\textbf{e}), the spectrum exhibits a symmetric Lorentzian lineshape.
A high-extinction Lorentzian profile is also observed when a zero is in the close vicinity of the unit circle so that its impact dominates, corresponding to the TZ condition (Fig.~\ref{fig3}\textbf{d}).
In contrast, when the pole and zero are located at comparable radial distances from the unit circle but separated in angle, their interference produces an asymmetric Fano lineshape.
This results in characteristic lineshapes with comparable peak and dip magnitudes.
The direction of this asymmetry is determined by the angular order of the pole and zero; in Fig.~\ref{fig3}\textbf{f}, the zero precedes the pole in counter-clockwise rotation, leading to a dip followed by a peak as wavelength increases, whereas in Fig.~\ref{fig3}\textbf{g}, the reversed angular ordering produces the opposite sequence.
These lineshape characteristics can also be interpreted as interference between resonant and non-resonant pathways with phase difference within the resonator system (see Section 3 of Supplementary Information) \cite{Fan:03,Yoon:13,Limonov2017}.

\section{Experimental Results}
%%%%%%%%%%%%%%%%%%%%%%%%%%%%%%%%%%%%%%%%%%%%%%%%%%%%%%%%%%%%%%%%%%%%%%%%%%%%%%%%%%%%%%%%%%%%%%%
\begin{figure*}[!th]
	\centering
	\includegraphics[width=0.95\linewidth]{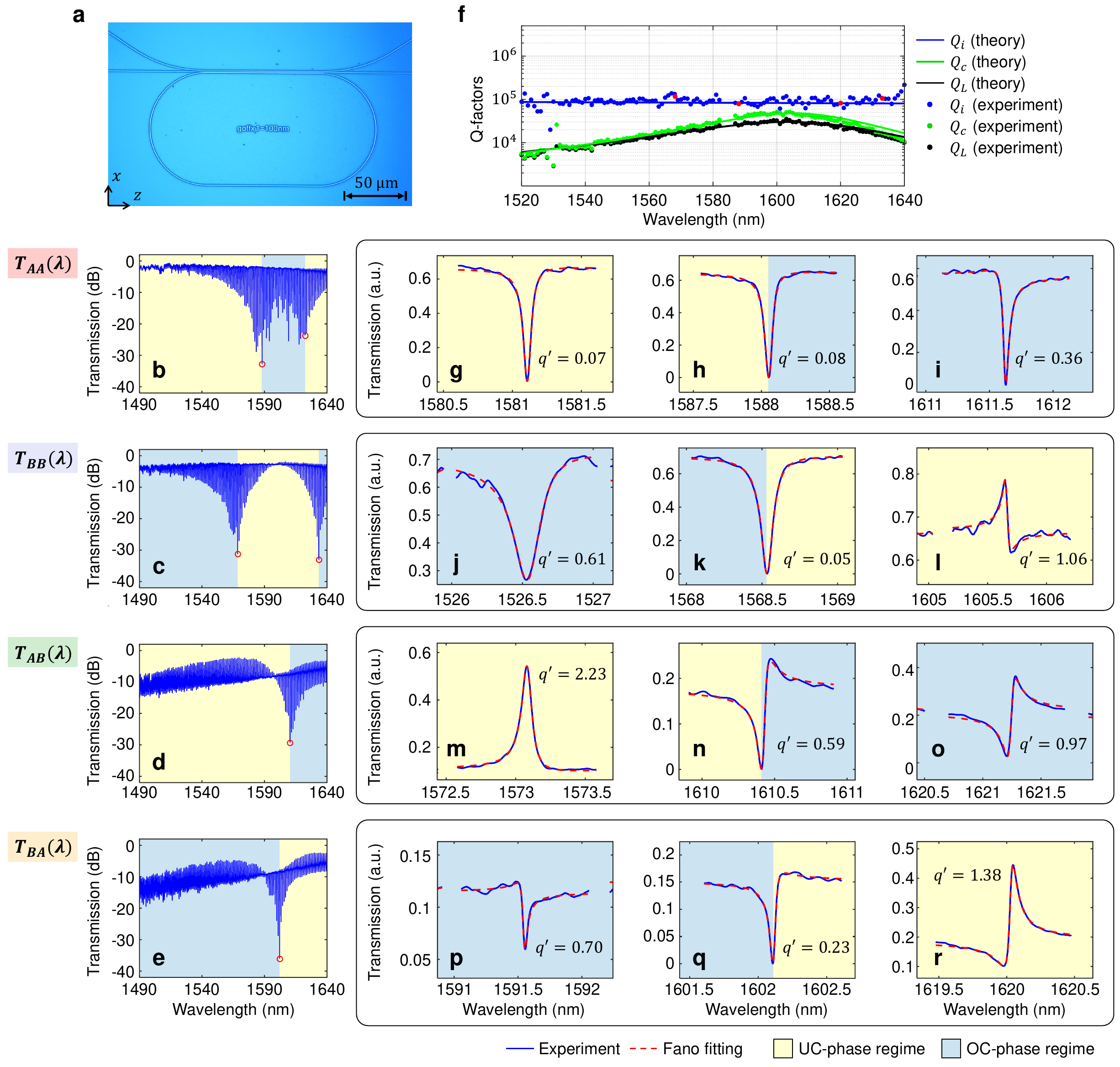}
	\caption{\textbf{Experimental characterization of the dual-bus resonator.}
        \textbf{a}, Optical microscope image of the fabricated device.
        \textbf{b--e}, Measured transmission spectra \textbf{b} $T_{AA}$, \textbf{c} $T_{BB}$, \textbf{d} $T_{AB}$, and \textbf{e} $T_{BA}$. 
        \textbf{f}, Extracted loaded ($Q_L$, black dots), intrinsic ($Q_i$, blue dots), and coupling ($Q_c$, green dots) quality factors; solid lines denote theoretical values.
        \textbf{g--r}, Representative resonance spectra (blue lines) in UC-phase, TZ, and OC-phase regimes: 
        \textbf{g--i} $T_{AA}$, \textbf{j--l} $T_{BB}$, \textbf{m--o} $T_{AB}$, and \textbf{p--r} $T_{BA}$. 
        Blue and yellow shaded regions represent OC- and UC-phase regimes, respectively.
        Each resonance is fitted to the generalized Fano model in Eq.~(\ref{eq:Fanofunction}) (red dashed lines), quantifying the Fano asymmetry parameter $q'=\sqrt{q^2+r^2}$ (annotated in each plot).
    }
	\label{fig4}
    % \vspace{1mm}
\end{figure*}

%%%%%%%%%%%%%%%%%%%%%%%%%%%%%%%%%%%%%%%%%%%%%%%%%%%%%%%%%%%%%%%%%%%%%%%%%%%%%%%%%%%%%%%%%%%%%%%%

To validate our theoretical analysis, we fabricated the dual-bus racetrack resonator on a silicon-on-insulator platform and measured its transmission spectra. 
Figure~\ref{fig4}\textbf{a} shows an optical microscope image of the fabricated device.
Detailed design parameters of the coupling gaps are provided in Section 1 of the Supplementary Information. 
Unless otherwise specified, the experimental results correspond to the same device geometries used in the theoretical simulations described above.

\subsection{Transmission spectra and coupling regimes}
Figures~\ref{fig4}\textbf{b}--\ref{fig4}\textbf{e} show the measured spectra $T_{AA}$, $T_{BB}$, $T_{AB}$, and $T_{BA}$, respectively, which closely match the simulated spectra in Figs.~\ref{fig2}\textbf{c}--\ref{fig2}\textbf{f}.
For $T_{AA}$, two resonances at 1588 nm and 1620 nm (red markers in Fig.~\ref{fig4}\textbf{b}) exhibit high extinction ratios, at the TZ condition of $\alpha=\left|{t_{11}}/{t_{22}}\right|$.
Similarly, for $T_{BB}$, high extinction resonances appear at 1568 nm and 1633 nm (red markers in Fig.~\ref{fig4}\textbf{c}). 
As in simulations, the two transmission zero points of $T_{AA}$ lie between those of $T_{BB}$.  
Comparison with Fig.~\ref{fig2} identifies complementary UC- and OC-phase regimes between $T_{AA}$ and $T_{BB}$, which is the key difference between the two.

For the measured cross-port spectra $T_{AB}$ and $T_{BA}$ (Figs.~\ref{fig4}\textbf{d} and \ref{fig4}\textbf{e}), transmission zero is observed at 1610 nm and 1602 nm, respectively, showing the highest extinction ratios.
%The spectral contours of $T_{AB}$ and $T_{BA}$ are similar in shape; however, as observed in simulations, they exhibit complementary coupling regimes with distinct phase responses across wavelength, determined by the zero locations of the device response.
Although the spectral trends of $T_{AB}$ and $T_{BA}$ appear similar in their lineshape and magnitude, they fundamentally differ in their coupling properties.
Specifically, they operate in complementary coupling regimes with different phase shifts, as visualized by the background colors in Figs.~\ref{fig4}\textbf{d} and \ref{fig4}\textbf{e}.
This complementary property is supported by our simulation in Figs.~\ref{fig2}\textbf{e} and \ref{fig2}\textbf{f} in connection with the pole-zero interpretation in Fig.~\ref{fig3}.

\subsection{Q-factor extraction and characterization}
For a conventional single-bus resonator, the $Q_L$ can be extracted from the resonance linewidth, while the intrinsic ($Q_i$) and coupling ($Q_c$) quality factors can be estimated using the minimum transmission dip at resonance ($T_{\mathrm{min}}$) \cite{Luo:11}.
A similar method can be applied to the dual-bus racetrack resonator. 
In analogy with the definition of $Q_L$ in \eqref{eq:totallossandQ}, the intrinsic and coupling quality factors are given by:
\begin{subequations}
\begin{align}
Q_i 
&=
\frac{\omega_0}{2\gamma_i} = \frac{\pi n_g L_{\mathrm{rt}}}{-\lambda_0 \ln|\alpha|},
\label{intrinsicQ} \\
Q_c 
&=
\frac{\omega_0}{2\gamma_c} = \frac{\pi n_g L_{\mathrm{rt}}}{-\lambda_0 \ln|t_{33}|},
\label{eq:couplingQ}
\end{align}
\end{subequations}
where $\lambda_0$ is the resonance wavelength, $n_g$ is the group index. Here $\gamma_i$ is the intrinsic amplitude decay rate related to the intrinsic loss $\alpha$ of the resonator and $\gamma_c$ is the coupling amplitude decay rate accounting for the total coupling loss due to the two bus waveguides.

To extract $Q_L$ from measured spectra, we fit each resonance to a general Fano resonance function \cite{Yoon:13,PhysRevB.87.125118}:
\begin{equation}
T(\lambda) =  p^2 \frac{\left( q + \dfrac{2(\lambda - \lambda_0)}{\gamma_L} \right)^2+r^2}{1 + \dfrac{4(\lambda - \lambda_0)^2}{\gamma_L^2}},
\label{eq:Fanofunction}
\end{equation}
where $p^2$ indicates the non-resonant intensity, $q$ the Fano asymmetry factor, $r$ another lineshape factor that accounts for the ratio between Fano and Lorentzian, and $\gamma_L$ the resonance linewidth. 
Since all four transfer functions share the same pole (i.e., the same $\gamma_L$), the extracted $Q_L$ is independent of the chosen spectrum. 
Figure~\ref{fig4}$\textbf{f}$ shows good agreement between the theoretically calculated and experimentally extracted $Q_L$ from $T_{AA}$ (Fig.~\ref{fig4}\textbf{b}) across the measured wavelength range.

For extracting $Q_i$, both $T_{AA}$ and $T_{BB}$ are required, since the parameters $t_{11}$, $t_{22}$, $t_{33}$, and $\alpha$ are related through their influence on both spectra.
Thus, it can be solved using four measurable quantities: the resonance dips $T_{AA, \mathrm{min}}$ and $T_{BB, \mathrm{min}}$, the non-resonant baseline intensity of $T_{AA}$, and the linewidth $\gamma_L$. 
As in the single-bus case, the correct sign of \(\pm\sqrt{T_{AA,\mathrm{min}}}\) and \(\pm\sqrt{T_{BB,\mathrm{min}}}\) must be selected according to the channel-defined coupling regime (either UC- or OC-phase regime). 
By consistently applying this procedure, we extract a nearly constant value of $\alpha \approx 0.95$ over the measured wavelength range, in close agreement with theory (Fig.~\ref{fig4}\textbf{f}).
The corresponding spectral behavior of \(|t_{11}(\lambda)|\), \(|t_{22}(\lambda)|\), and \(|t_{33}(\lambda)|\) is provided in Section 5 of Supplementary Information.

\subsection{Lorentzian and Fano resonance lineshapes}
The spectral response of each port can be further understood through its resonance lineshape.
While symmetric Lorentzian profiles arise from isolated resonant modes, asymmetric Fano resonances originate from interference between resonant and non-resonant pathways, described by the Fano parameter $q$ \cite{Yoon:13,Limonov2017}. 
To correctly quantify the relative strength of the resonant component compared to the non-resonant background, we adopt the generalized Fano asymmetry parameter $q' = \sqrt{q^2 + r^2}$ from the general Fano model in \eqref{eq:Fanofunction}  \cite{PhysRevB.87.125118}.

The measured resonance lineshapes of $T_{AA}$ (Figs.~\ref{fig4}\textbf{g}--\ref{fig4}\textbf{i}) exhibit nearly zero $q'$, corresponding to ideal Lorentzian dips across the entire wavelength range. 
For $T_{BB}$ (Figs.~\ref{fig4}\textbf{j}--\ref{fig4}\textbf{l}), the resonance mostly exhibits Lorentzian lineshapes ($q' \approx 0$), except near 1605 nm, where a Fano-like asymmetry with $q' \approx 1$ appears. 
This occurs when the pole and zero lie at nearly the same radial distance but differ in angular separation, resulting in a Fano resonance. 
Resonances in this region exhibit a significantly reduced extinction ratio, consistent with the condition $\left|{t_{22}}/{t_{11}}\right| \gg \alpha$, corresponding to a zero far from the unit circle, in the highly UC-phase regime.

For the cross-port spectra $T_{AB}$ and $T_{BA}$, lineshape variations are more pronounced due to the wavelength-dependent evolution of relative pole-zero positions. 
For $T_{AB}$, Lorentzian peaks ($q'\gg1$) in the UC-phase regime (Fig.~\ref{fig4}\textbf{m}) evolve into highly asymmetric Fano resonances with $\approx$20~dB extinction at transmission zero (Fig.~\ref{fig4}\textbf{n}), and further into Fano resonances with $q' \approx 1$ in the OC-phase regime (Fig.~\ref{fig4}\textbf{o}).
Even more diverse behavior is observed in $T_{BA}$, ranging from Lorentzian peaks in the highly OC-phase regime to Fano resonances with both negative and positive $q$ values (Figs.~\ref{fig4}\textbf{p}, \ref{fig4}\textbf{r}). 
Intermediate regimes also exhibit near-symmetric dips (Fig.~\ref{fig4}\textbf{q}).
Notably, both extinction ratio and resonance asymmetry can be systematically engineered through device geometry. 
For instance, varying the additional gap offsets in $g_2(z)$ shifts the spectral position of the transmission zero: increasing the offset induces a red shift, whereas decreasing it causes a blue shift (see Section 4 of Supplementary Information).

\section{Discussion}

\normalsize{}
%%%%%%%%%%%%%%%%%%%%%%%%%%%%%%%%%%%%%%%%%%%%%%%%%
\begin{figure}[!t]
\centering
\includegraphics[width=0.95\columnwidth]{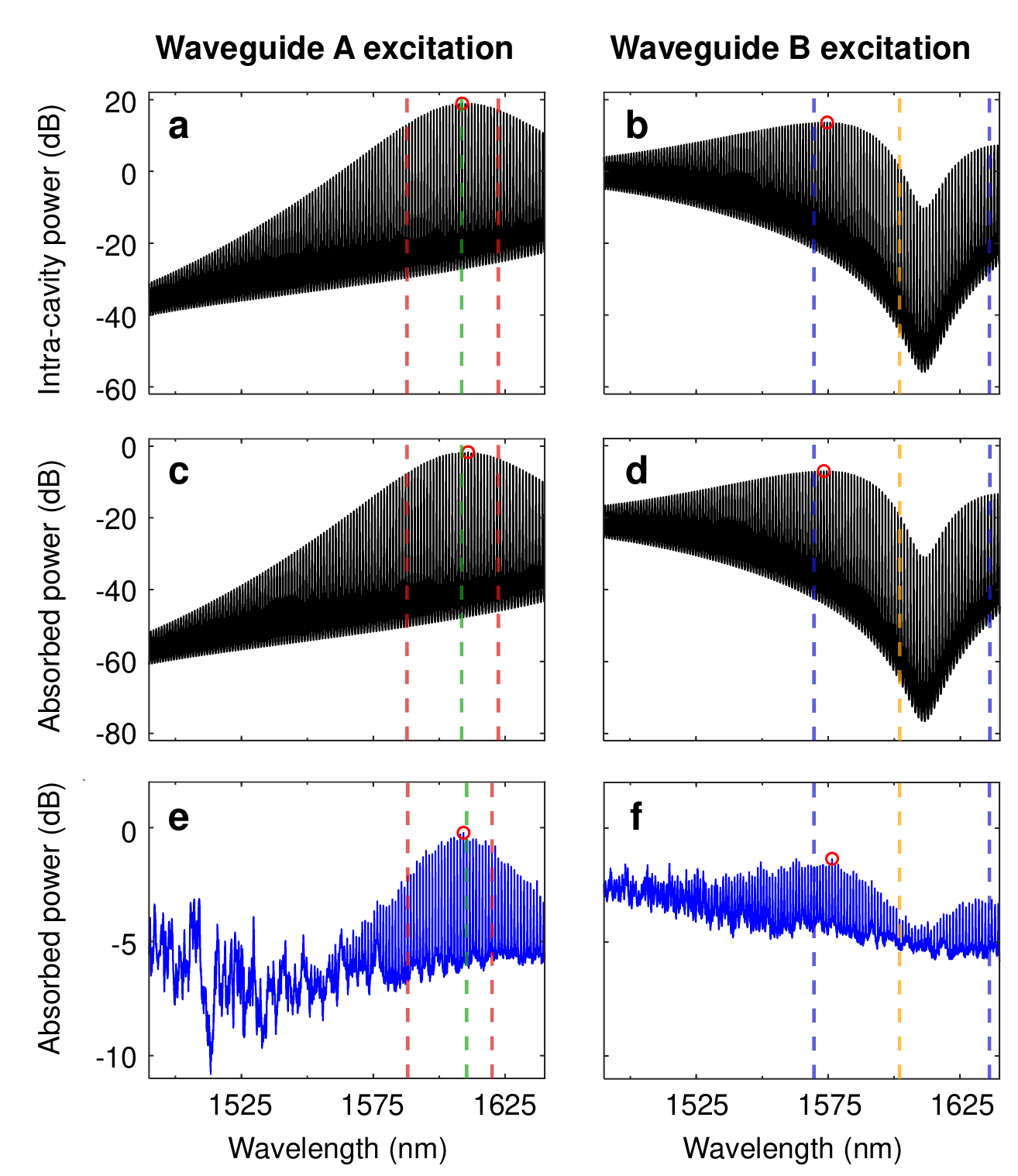}
    \caption{\textbf{Intra-cavity power and transmission zeros.} 
    \textbf{a}, \textbf{b} Simulated intra-cavity powers $P_{\text{cavity},A}$ and $P_{\text{cavity},B}$ for waveguide excitations $A$ and $B$, respectively. 
    \textbf{c}, \textbf{d}, Simulated absorbed powers $P_{\text{abs},A}$ and $P_{\text{abs},B}$. 
    \textbf{e}, \textbf{f}, Corresponding experimental results of absorbed power. 
    Red circles mark the wavelengths of maximum power. 
    Vertical colored lines indicate the transmission-zero wavelengths for $T_{AA}$ (red), $T_{BB}$ (blue), $T_{AB}$ (green), and $T_{BA}$ (yellow).
    }
\label{fig5}
% \vspace{1mm}
\end{figure}
%%%%%%%%%%%%%%%%%%%%%%%%%%%%%%%%%%%%%%%%%%%%%%%%%
Our analysis of the dual-bus configuration reveals broader implications for the physics of multi-port resonators, showing that they exhibit coupling behavior fundamentally distinct from conventional single-bus systems.
From the extracted quality factors (Fig.~\ref{fig4}\textbf{f}), the $Q_c$ is consistently lower than the $Q_i$ across the measured spectrum. 
This indicates that the resonator operates in a {globally over-coupled regime}, where the total coupling loss rate exceeds the intrinsic loss rate ($\gamma_c > \gamma_i$). 
In a single-bus resonator, such an over-coupled state would normally preclude the appearance of a transmission zero. 
However, our device exhibits deep extinction notches for all four transmission channels ($T_{AA}$, $T_{BB}$, $T_{AB}$, and $T_{BA}$), demonstrating that the condition for a channel-specific transmission zero is {decoupled from the resonator's intrinsic power balance}, as explained below in detail.

This decoupling becomes evident when analyzing the intra-cavity power, a key metric for light-matter interaction. 
The intra-cavity powers for excitation from waveguides $A$ and $B$ ($P_{\text{cavity},A}$ and $P_{\text{cavity},B}$) are expressed as:
%%%%%%%%%%%%%%%%%%%%%%%%%%%%%%%%%%%%%%%%%%%
\begin{subequations}
\begin{align}
P_{\text{cavity},A} 
&=
\left| \frac{t_{31} \alpha e^{i\theta}}{1-t_{33}^* \alpha e^{i\theta}} \right|^2,
\label{intracavityA} \\
P_{\text{cavity},B} 
&=
\left| \frac{t_{32} \alpha e^{i\theta}}{1-t_{33}^* \alpha e^{i\theta}} \right|^2.
\label{intracavityB}
\end{align}
\end{subequations}
%%%%%%%%%%%%%%%%%%%%%%%%%%%%%%%%%%%%%%%%%%%
Figures~\ref{fig5}\textbf{a} and \ref{fig5}\textbf{b} show the simulated intra-cavity powers for excitation from waveguides $A$ and $B$, respectively, highlighting the clear mismatch between the wavelengths of maximum intra-cavity power and those of transmission zeros. 
Dashed lines mark the TZ condition points from Fig.~\ref{fig2}, while red circles denote the intra-cavity power maxima. 
Notably, none of the transmission-zero wavelengths coincide with the wavelength of maximum stored power. 
This behavior can be confirmed experimentally by the absorbed power, defined as $P_{\text{abs},A}=1-T_{AA}-T_{AB}$ and $P_{\text{abs},B}=1-T_{BB}-T_{BA}$.
Figures~\ref{fig5}\textbf{c} and \ref{fig5}\textbf{d} show the simulated absorbed powers, while Figs.~\ref{fig5}\textbf{e} and \ref{fig5}\textbf{f} are the corresponding experimental results from Figs.~\ref{fig4}.
The excellent agreement between simulation and experiment confirms that maximizing stored power and achieving zero transmission does not necessarily coincide as opposed to conventional interpretations of single-bus resonators.

These findings separate the conventional notion of critical coupling into three physically independent conditions in a multi-port system:
\begin{enumerate}
    \item \textbf{Critical coupling} ($\gamma_i=\gamma_c$): A cavity-defined condition where intrinsic and external decay rates are equal ($Q_i=Q_c$). In a single-bus resonator, this condition coincides with a transmission zero, but in multi-port configurations, this coincidence generally breaks down.
    \item \textbf{Transmission zero} ($T_{ij} = 0$): An interference-based condition specific to a given input-output path, occurring when direct and resonator-mediated fields destructively interfere.
    \item \textbf{Maximum intra-cavity power} ($P_{\text{cavity}}^{\max}$): An excitation-dependent condition that determines peak stored energy, critical for enhancing light-matter interactions for nonlinear, laser, and sensing processes.
\end{enumerate}
This fundamental separation of conditions necessitates terms to describe the resonator's interference behavior.
To this end, we define UC-phase and OC-phase regimes based on the phase shift of the resonance dips. 
This classification, analogous to minimum-phase and maximum-phase systems, provides the necessary framework to fully account for the channel-specific transmission in multi-port resonators.
Moreover, this separation shows that conditions traditionally unified in single-bus resonators can be independently controlled in a multi-port system, introducing a new design paradigm for resonators.
It enables different functionalities to be realized at different wavelengths--for example, achieving a transmission zero for filtering at one wavelength while maximizing intra-cavity power for nonlinear conversion at another.
Such independent control provides new opportunities to overcome the inherent limitations of single-bus resonators and to tailor device performance for specific application. 

It is also important to note that the proposed scheme is not restricted to silicon. 
Beyond the silicon platform, we have also confirmed the versatility of this coupling scheme on other platforms such as \SiN, which offers a wide transparency window and low propagation loss.
This makes it highly suitable for broadband and nonlinear applications, including frequency conversion. 
Simulations on \SiN~show broadband operation over 500–1800~nm, and experimental results confirm that the key features--complementary coupling responses, Fano resonances, and the separation of coupling conditions--remain preserved (see Section 6 of Supplementary Information). 
These results confirm that the proposed dual-bus scheme is readily applicable across various material platforms and wavelength ranges, providing a generalized framework for multi-port spectral engineering in integrated photonics. 

\section{Conclusion}

\normalsize{}
In conclusion, we have introduced and experimentally validated a dual-bus racetrack resonator that enables broadband dispersive coupling with channel-specific complementary spectral responses. 
This scheme decouples the conditions for transmission zero, critical coupling, and maximum intra-cavity power---traditionally indistinguishable in single-bus resonators---thereby establishing a new design paradigm for multi-port bus-resonator coupling systems.  
We have developed a comprehensive model based on coupled-mode theory and pole-zero analysis, which accurately describes the spectral responses and provides clear physical insight into the observed phenomena. 
The model also refines the concepts of UC- and OC-phase regimes according to the transfer function's phase response, distinguishing them from classical coupling-regime definitions.
In addition, the dual-bus configuration inherently generates wavelength-dependent Fano resonances, arising from engineered phase differences between the system's poles and zeros. 
Overall, these results establish a versatile framework for multi-port spectral engineering in microresonators, paving the way for reconfigurable filters, high-speed modulators, multi-spectral resonators, sensors, and advanced nonlinear applications.
%%%%%%%%%%%%%%%%%%%%%%%%%%%%%%%%%%%%%%%%%%%%%%%%%%%%%%%%%%%%%%%%%%%%%%%%%%%%%%%%%%%%%%%%%%%%%%%%%

\section*{Acknowledgments}
This work was supported by the National Research Foundation of Korea (NRF) funded by the Korea government (MSIT) (RS-2023-00210997).
It was also partially supported by the Asian Office of Aerospace Research and Development (FA23862514056) and the KAIST Undergraduate Research Participation program.\\

\section*{Author contributions} 
TK developed the theoretical framework, performed device characterization, and analyzed the results in support of MH.
MH fabricated the device, and YC contributed to the device design.
JY and SK supervised the project.
TK and SK prepared the manuscript, and all authors discussed the results and contributed to the final version.\\

\section*{Competing interests}
The authors declare no competing financial interests.

% \section*{Data Availability Statement} 

\bibliographystyle{naturemag}

\clearpage
\onecolumngrid

\begin{center}
\textbf{\large Supplementary Information for: \\Dual-Bus Resonator for Multi-Port Spectral Engineering}
\end{center}

% Reset all counters
\setcounter{page}{1}
\setcounter{section}{0}
\setcounter{figure}{0}
\setcounter{table}{0}
\setcounter{equation}{0}

% Format new counters to be S1, S2, etc.
\renewcommand{\thepage}{S\arabic{page}}
\renewcommand{\thesection}{S\arabic{section}}
\renewcommand{\thefigure}{S\arabic{figure}}
\renewcommand{\thetable}{S\arabic{table}}
\renewcommand{\theequation}{S\arabic{equation}}

\section{Device design and fabricaion}

%%%%%%%%%%%%%%%%%%%%%%%%%%%%%%%%%%%%%%%%%%
\begin{figure*}[htbp]
  \centering
  \includegraphics[width=0.98\linewidth]{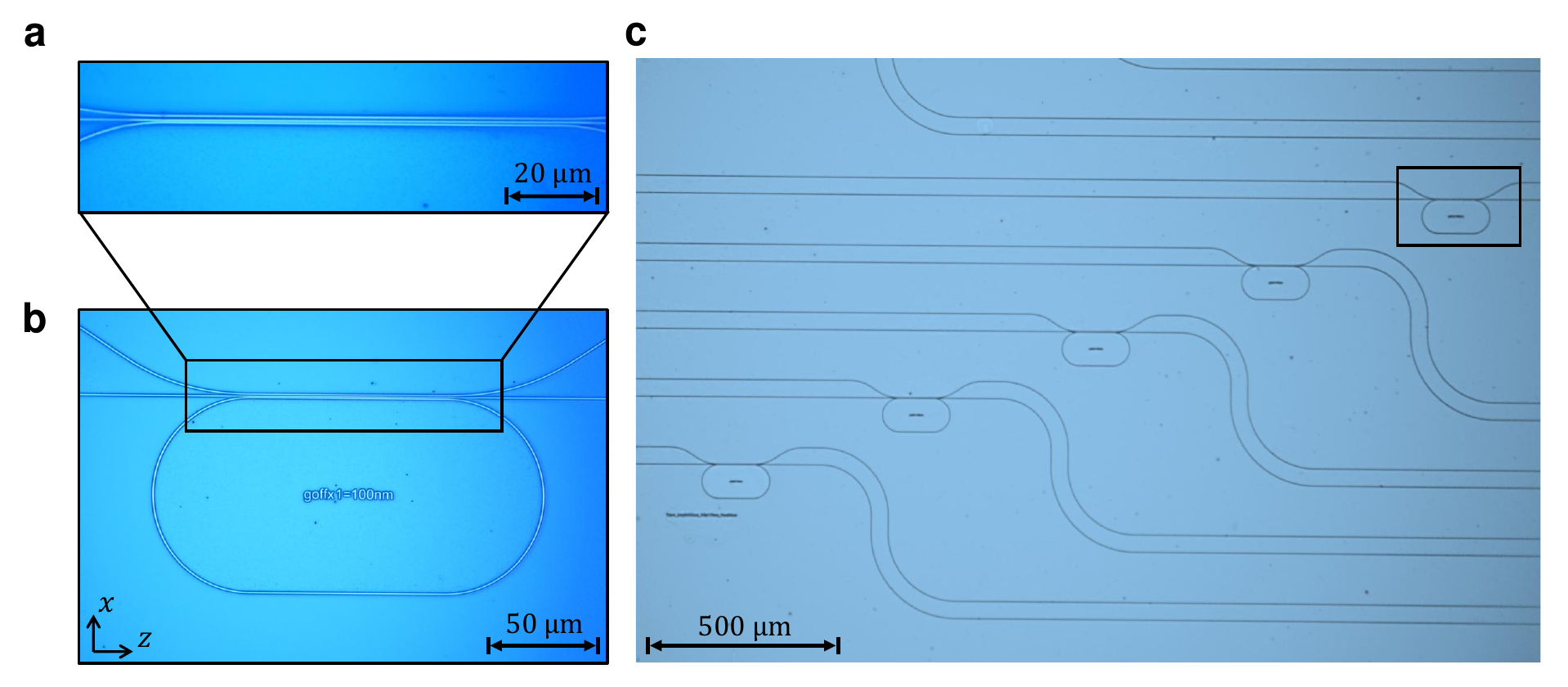}
  \caption{\textbf{Optical microscope images of the fabricated dual-bus racetrack resonator.}
  \textbf{a}, \(100\times\) magnification of the three-waveguide coupler section.
  \textbf{b}, \(50\times\) magnification of a device with \(g_{1,\mathrm{off}}=\SI{0}{\nano\metre}\) and \(g_{2,\mathrm{off}}=\SI{100}{\nano\metre}\).
  \textbf{c}, Five devices with \(g_{2,\mathrm{off}}=\SIlist{0;25;50;75;100}{\nano\metre}\), arranged left to right.
  (Black box represents the device in \textbf{b}.)
  }
  \label{SIfig1}
  \vspace{1mm}
\end{figure*}
%%%%%%%%%%%%%%%%%%%%%%%%%%%%%%%%%%%%%%%%%%

All waveguides, including the resonator, share identical cross-sectional dimensions of width $w=\SI{500}{\nano\metre}$ and height $h=\SI{220}{\nano\metre}$.
These dimensions are designed to support only the fundamental transverse electric (TE) mode.
The radius of curvature of the racetrack resonator is $R=\SI{50}{\micro\metre}$.

Waveguide $B$ is designed with a hyperbolic cosine function bend, causing $g_1(z)$, the distance between waveguide $A$ and $B$, and $g_2(z)$, the distance between waveguide $B$ and $C$, to vary over the coupling region length $L=\SI{100}{\micro\metre}$. 
The gaps are designed as:

\begin{equation}
g_1(z) = g_{1,\mathrm{off}} + 0.3 
- \frac{1}{4} \cdot \frac{\cosh{\left[0.012(z - 190)\right]}}{\cosh{\left(0.012 \cdot 190\right)}} 
\quad \left[\si{\micro\meter}\right],
\label{eq:gap1}
\end{equation}

\begin{equation}
g_2(z) = g_{2,\mathrm{off}} + 0.3 
+ \frac{1}{4} \cdot \frac{\cosh{\left[0.012(z - 190)\right]}}{\cosh{\left(0.012 \cdot 190\right)}} 
\quad \left[\si{\micro\meter}\right],
\label{eq:gap2}
\end{equation}

where $g_{1,\mathrm{off}}$ and $g_{2,\mathrm{off}}$ are the uniform gap offsets in $g_1(z)$ and $g_2(z)$ respectively. 
The distance between waveguide $A$ and $B$ ($g_1(z)+g_2(z)$) is always uniform over the propagation distance $z$.
When $g_{1,\mathrm{off}}$ and $g_{1,\mathrm{off}}$ are both zero, $g_1(z)+g_2(z)$ is $\SI{600}{\nano\metre}$.
Due to the bending of waveguide $B$, the propagation constant $\beta_B$ is slightly different from that of waveguide $A$ and $C$ ($\beta_A=\beta_C=\beta$) even though all three waveguides share the same cross-sectional dimensions.
This difference in the propagation constant is denoted by $\Delta\beta=\beta_B-\beta$.

The device is fabricated on a standard silicon-on-insulator (SOI) wafer with a \SI{220}{\nano\metre} silicon device layer and a \SI{2}{\micro\metre} buried oxide (BOX) layer.
The device layout is patterned using electron beam lithography (JEOL system, 100~keV) with hydrogen silsesquioxane (HSQ) resist. 
The pattern is then transferred onto the silicon layer through dry etching using a Cl\textsubscript{2}/O\textsubscript{2} plasma chemistry. 
After etching, a \SI{2}{\micro\metre} PECVD-deposited SiO\textsubscript{2} upper cladding is applied to provide optical confinement and structural protection.

Figure~\ref{SIfig1}(a) shows an optical microscope image of the fabricated device under 100$\times$ magnification, highlighting the dual-bus waveguides with varying coupling gaps along the interaction region. 
Figure~\ref{SIfig1}(b) presents a set of five fabricated devices, where the offset gap parameter $g_{2,\mathrm{off}}$  varies across 0 nm, 25 nm, 50 nm, 75 nm, and 100 nm, while $g_{1,\mathrm{off}}$ is fixed at 0 nm. 
This systematic variation enables a comparative study of the spectral behavior induced by the gap distance. 
Unless otherwise noted, the theoretical simulations and experimental results in the paper correspond to the device with $g_{1,\mathrm{off}} = \SI{0}{\nano\metre}$ and $g_{2,\mathrm{off}} = \SI{100}{\nano\metre}$.

\section{Modeling of the three-waveguide coupler}

We solve the coupled-mode equation for the lossless three-waveguide directional coupler~\cite{PASPALAKIS200630} to find the transfer matrix $\mathbf{T}$ of the coupling region.
It is a $3\times3$ para-unitary matrix with elements $t_{ij}$ denoting the transmission coefficients, written as follows:
%%%%%%%%%%%%%%%%%%%%%%%%%%%
\begin{equation}
\begin{pmatrix}
A(L) \\
B(L) \\
C(L)
\end{pmatrix}
=
\begin{pmatrix}
t_{11} & t_{12} & t_{13} \\
t_{21} & t_{22} & t_{23} \\
t_{31} & t_{32} & t_{33}
\end{pmatrix}
\begin{pmatrix}
A(0) \\
B(0) \\
C(0)
\end{pmatrix}.
\label{eq:TMM}
\end{equation}
%%%%%%%%%%%%%%%%%%%%%%%%%%%

Four transfer functions $t_{AA}$, $t_{BB}$, $t_{AB}$, and $t_{BA}$ are defined based on its input-output configurations as: 
 %%%%%%%%%%%%%%%%%%%%%%%%%%%%%%%%%%%%%%%%%%%%%%%%%%%%%%%%%
 \begin{subequations}
 \begin{align}
 t_{AA}
 &=
 \frac{A(L)}{A(0)}
 =
 t_{11}+\frac{t_{13} t_{31} \alpha e^{i \theta}}{1-t_{33} \alpha e^{i\theta}},
 \label{eq:generalHAA} \\
 t_{BB}
 &=
 \frac{B(L)}{B(0)}
 =
 t_{22}+\frac{t_{23} t_{32} \alpha e^{i \theta}}{1-t_{33} \alpha e^{i\theta}},
 \label{eq:generalHBB} \\
 t_{AB}
 &=
 \frac{B(L)}{A(0)}
 =
 t_{21}+\frac{t_{23} t_{31} \alpha e^{i \theta}}{1-t_{33} \alpha e^{i\theta}},
 \label{eq:generalHAB} \\
 t_{BA}
 &=
 \frac{A(L)}{B(0)}
 =
 t_{12}+\frac{t_{13} t_{32} \alpha e^{i \theta}}{1-t_{33} \alpha e^{i\theta}},
 \label{eq:generalHBA}
 \end{align}
 \end{subequations}
 %%%%%%%%%%%%%%%%%%%%%%%%%%%%%%%%%%%%%%%%%%%%%%%%%%%%%%%%%
 where $\alpha$ is the round-trip loss coefficient (field amplitude fraction) and $\theta$ is the corresponding accumulated round-trip phase of the racetrack resonator. 
 The closed-loop relation $C(0)=C(L)\alpha e^{i\theta}$ is used to derive the above expressions.

 The equations~\eqref{eq:generalHAA}-~\eqref{eq:generalHBA} can be further simplified using the para-unitary condition of $\mathbf{T}$. Since $\mathbf{T^{-1}}=\mathbf{T^{\dagger}}$ and $\mathbf{T^{-1}}=\frac{\text{adj}\,\mathbf{T}}{\det{\mathbf{T}}}$, comparing $\mathbf{T^{-1}}$ element by element, we have:
 %%%%%%%%%%%%%%%%%%%%%%%%%%%%%%%%%%%%%%%%%%%
 \begin{align}
 t_{11}^* e^{i\varphi} &= t_{22}t_{33} - t_{13}t_{31}, \nonumber \\
 t_{22}^* e^{i\varphi} &= t_{11}t_{33} - t_{23}t_{32}, \nonumber \\
 t_{12}^* e^{i\varphi} &= t_{21}t_{33} - t_{23}t_{31}, \nonumber \\
 t_{21}^* e^{i\varphi} &= t_{12}t_{33} - t_{13}t_{32},
 \end{align}
 \label{eq:matrixrelation}
 %%%%%%%%%%%%%%%%%%%%%%%%%%%%%%%%%%%%%%%%%%
 where $\varphi$ is the global phase of the three-waveguide coupler, which signifies a common phase shift experienced by light as it propagates through the coupling region.

 Thus, the transfer functions become the following:
 %%%%%%%%%%%%%%%%%%%%%%%%%%%%%%%%%%%%%%%%%%%%%%%
 \begin{subequations}
 \begin{align}
 t_{AA}
 &=
 \frac{t_{11}- t_{22}^* e^{i\varphi} \alpha e^{i \theta}}{1-t_{33} \alpha e^{i\theta}},
 \label{eq:simpleHAA} \\
 t_{BB}
 &=
 \frac{t_{22}- t_{11}^*e^{i\varphi} \alpha e^{i \theta}}{1-t_{33} \alpha e^{i\theta}},
 \label{eq:simpleHBB} \\
 t_{AB}
 &=
 \frac{t_{21}- t_{12}^*e^{i\varphi} \alpha e^{i \theta}}{1-t_{33} \alpha e^{i\theta}},
 \label{eq:simpleHAB} \\
 t_{BA}
 &=
 \frac{t_{12}- t_{21}^*e^{i\varphi} \alpha e^{i \theta}}{1-t_{33} \alpha e^{i\theta}}.
 \label{eq:simpleHBA}
 \end{align}
 \end{subequations}
 %%%%%%%%%%%%%%%%%%%%%%%%%%%%%%%%%%%%%%%%%%%%%%%%%%%

\newpage
\section{Phase difference and Fano resonance}

%%%%%%%%%%%%%%%%%%%%%%%%%%%%%%%%%%%%%%%%%%
\begin{figure*}[htbp]
  \centering
  \includegraphics[width=0.98\linewidth]{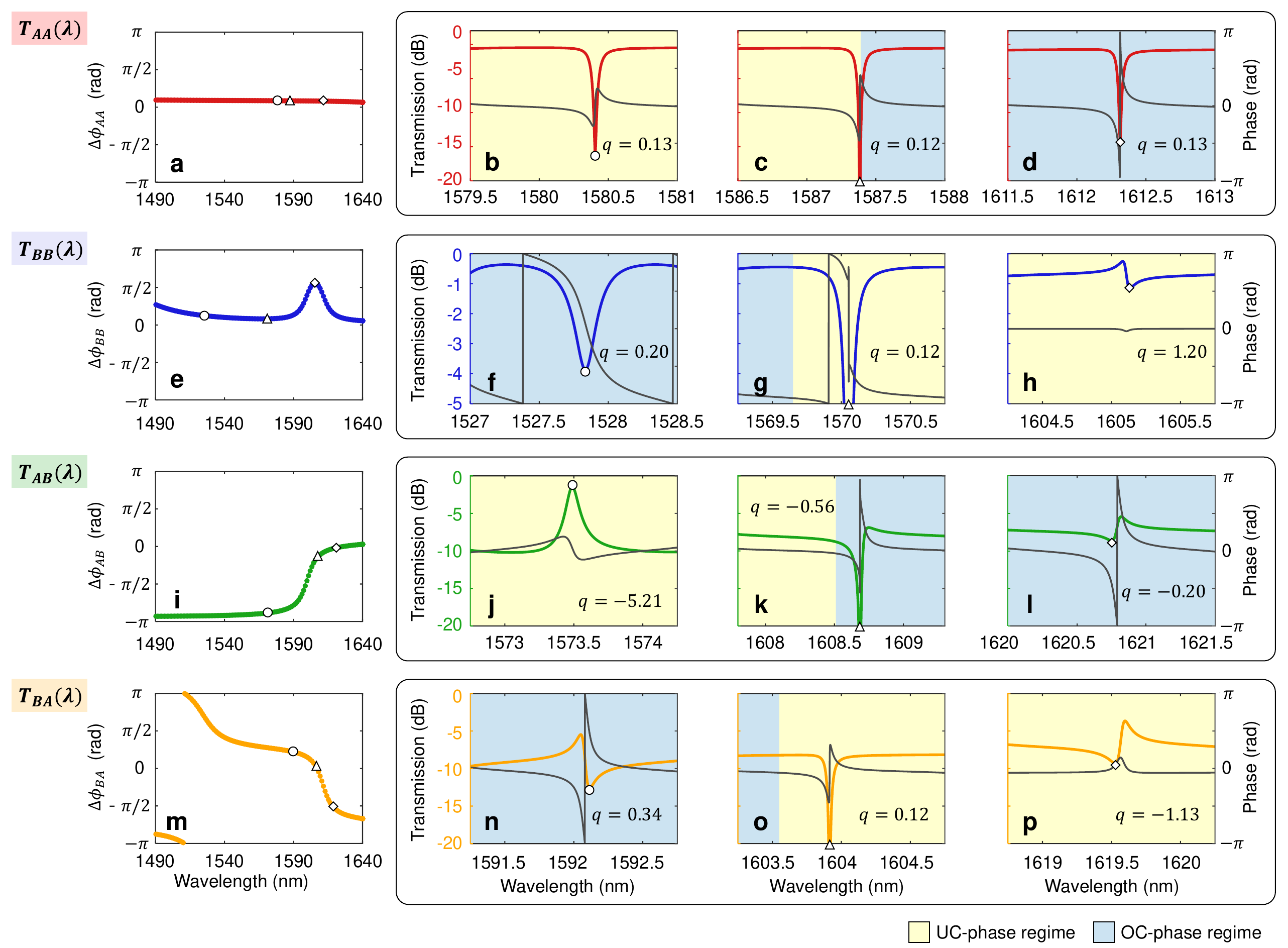}
  \caption{\textbf{Phase differences and resonance lineshapes in the four transmission spectra.}
  \textbf{a}, Phase difference \(\Delta\phi_{AA}\).
  \textbf{b--d}, Magnitude and phase of \(T_{AA}(\lambda)\) near resonance for three regimes:
  (b) port-specific under-coupling, (c) port-specific critical coupling, and (d) port-specific over-coupling.
  \textbf{e}, Phase difference \(\Delta\phi_{BB}\).
  \textbf{f--h}, Magnitude and phase of \(T_{BB}(\lambda)\) near resonance for:
  (f) port-specific over-coupling, (g) port-specific critical coupling, and (h) port-specific under-coupling.
  \textbf{i}, Phase difference \(\Delta\phi_{AB}\).
  \textbf{j--l}, Magnitude and phase of \(T_{AB}(\lambda)\) near resonance for:
  (j) port-specific under-coupling, (k) port-specific critical coupling, and (l) port-specific over-coupling.
  \textbf{m}, Phase difference \(\Delta\phi_{BA}\).
  \textbf{n--p}, Magnitude and phase of \(T_{BA}(\lambda)\) near resonance for:
  (n) port-specific over-coupling, (o) port-specific critical coupling, and (p) port-specific under-coupling.}
  \label{SIfig2}
  \vspace{1mm}
\end{figure*}
%%%%%%%%%%%%%%%%%%%%%%%%%%%%%%%%%%%%%
While the spectra can be analyzed rigorously via pole--zero theory, a two-path interference picture is often used to explain Fano resonance in photonics~\cite{Limonov2017}. 
In this view, each transmission \(T_{ij}(\lambda)\) results from the interference between a non-resonant (direct) pathway---the first term in Eqs.~\eqref{eq:generalHAA}--\eqref{eq:generalHBA}---and a resonant pathway that couples into the cavity and out after one or more round trips (second term).

The relative phase \(\Delta\phi\) between these two pathways sets the Fano asymmetry. With the convention used here, the Fano parameter is related to the phase by~\cite{Gu2020,Yoon:13}
\[
q=\tan\!\left(\frac{\Delta\phi}{2}\right).
\]
Thus, at wavelengths where \(\Delta\phi\!\approx\!0 \ (\mathrm{mod}\ 2\pi)\) the Fano asymmetry coefficient $q$ yields \(|q|\!\approx\!0\) (Lorentzian-like lineshape), \(\Delta\phi\!\approx\!\pm\pi\) yields \(q\!\rightarrow\!\pm\infty\) (symmetric peak), and \(\Delta\phi\!\approx\!\pm\pi/2\) gives \(|q|\!\approx\!1\) (maximally asymmetric Fano profile).
The calculated phase differences for all ports are shown in Figs.~\ref{SIfig2}\textbf{a}, \textbf{e}, \textbf{i}, and \textbf{m} ($\Delta\phi_{ij}$).

For the through ports, \(T_{AA}(\lambda)\) and \(T_{BB}(\lambda)\), \(\Delta\phi\) stays near \(0\) over broad bands, producing predominantly Lorentzian dips (\(|q|\approx0\)).
Local excursions of \(\Delta\phi_{BB}\) toward \(\pi/2\) near 1600 nm introduce visible asymmetry as shown in Figs.~\ref{SIfig2}\textbf{h} and the lineshape becomes strongly asymmetric with \(|q|\approx1\).

In contrast, the cross ports \(T_{AB}(\lambda)\) and \(T_{BA}(\lambda)\) exhibit wider \(\Delta\phi\) variations with wavelength, leading to a progression of lineshapes (Figs.~\ref{SIfig2}\textbf{j--l}, \textbf{n--p}). 
As \(\Delta\phi_{AB}\) sweeps from \(-\pi\) to \(0\), \(q\) evolves from near \(-\infty\) (symmetric peak) through \(|q|\approx-1\) (maximally asymmetric Fano) toward \(|q|\approx0\) (Lorentzian-like dip). 
The sign of \(q\) (set by \(\Delta\phi\)) determines whether the peak precedes or follows the resonance dip.

This interference picture is most accurate when the resonant and non-resonant amplitudes are comparable. In our dual-bus racetrack resonator, there are spectral regions where that condition breaks down (e.g., when the resonant pathway dominates), so interpreting \(q\) purely as a measure of resonant-to-background strength can be misleading at certain wavelengths~\cite{PhysRevB.87.125118}.

\newpage
 \section{Experimental results for various gap offsets}

 \begin{figure*}[h!]
    \centering
    \includegraphics[width=0.9\textwidth]{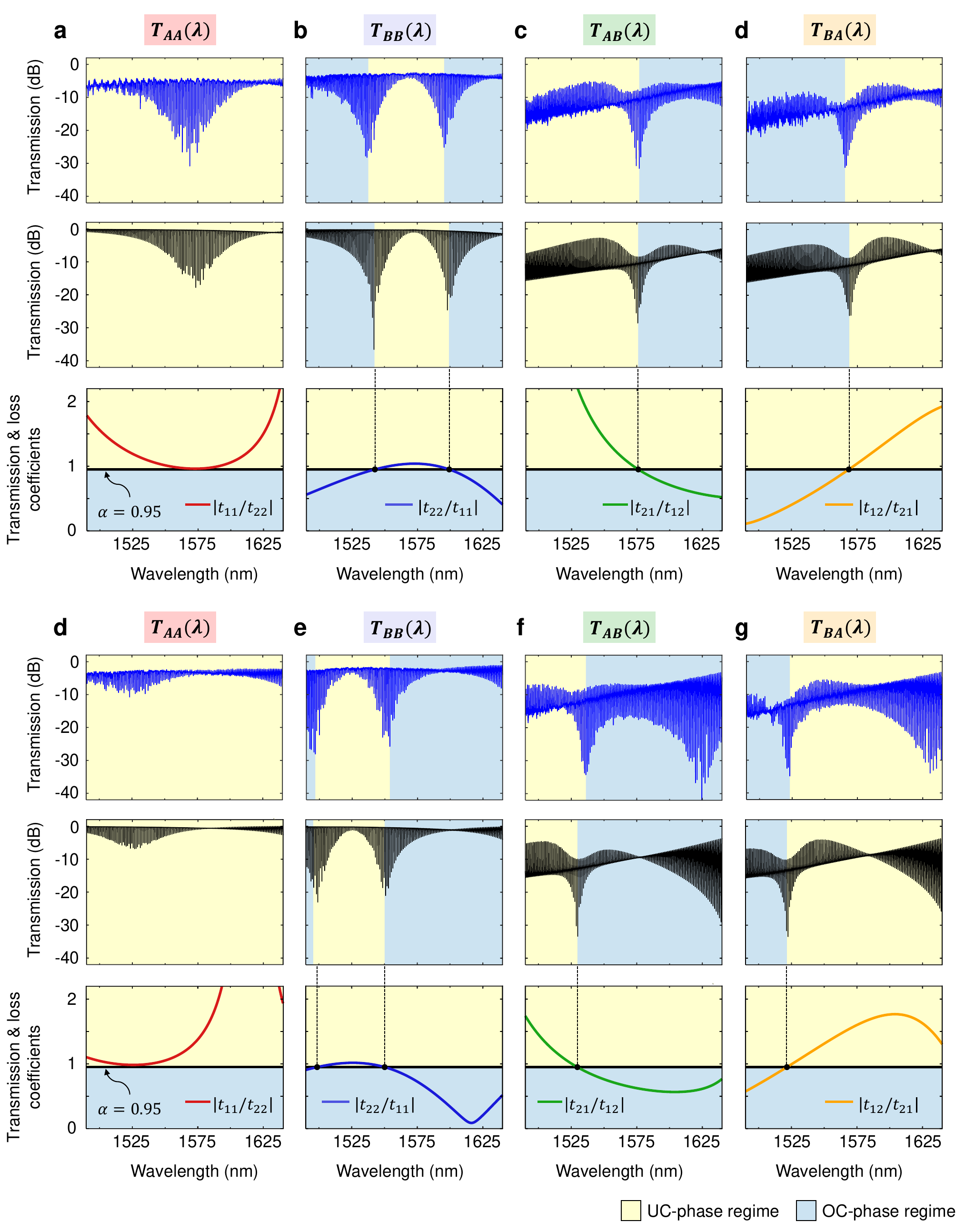}
    \caption{
    \textbf{Experimental and simulation of transmission spectra for device with offset $g_{2,\mathrm{off}}$.} 
    \textbf{a}-\textbf{d}, Device with $g_{1,\mathrm{off}}= \SI{0}{\nano\metre}$ and $g_{2,\mathrm{off}}= \SI{75}{\nano\metre}$.
    The top row of plots shows experimentally measured transmission spectra ($T_{AA}(\lambda)$, $T_{BB}$, $T_{AB}$, and $T_{BA}$ respectively) in blue, while the second row shows the corresponding simulated spectra.
    The final row shows the extracted transmission and loss coefficients.
    \textbf{e}-\textbf{h}, Device with $g_{1,\mathrm{off}}= \SI{0}{\nano\metre}$ and $g_{2,\mathrm{off}}= \SI{50}{\nano\metre}$. 
    The panels are organized in the same manner as in \textbf{a}-\textbf{d}.
    }
    \label{fig:SIfig3}
\end{figure*}

\begin{figure*}[h!]
    \centering
    \includegraphics[width=0.9\textwidth]{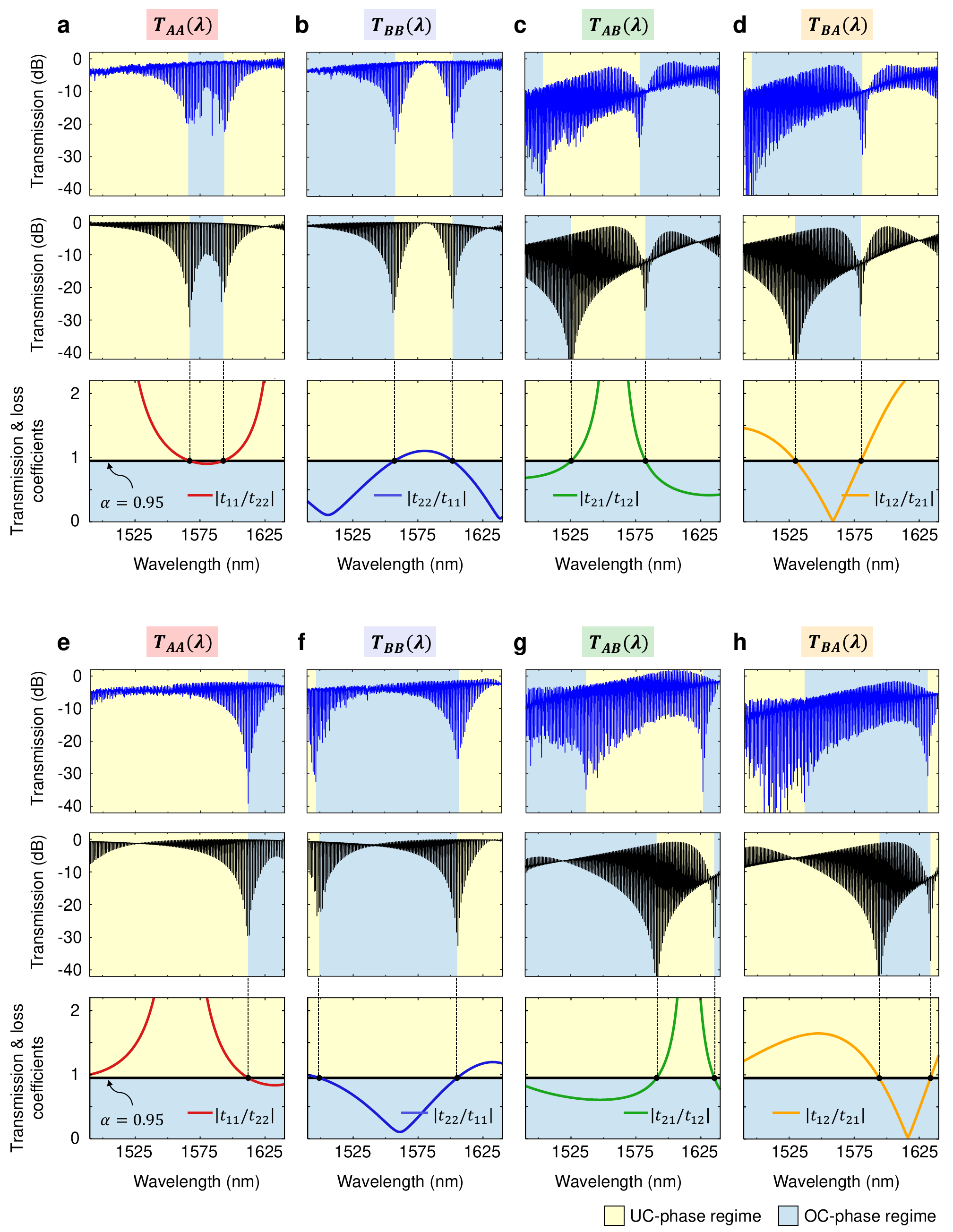}
    \caption{
    \textbf{Experimental and simulation of transmission spectra for device with offset $g_{1,\mathrm{off}}$.} 
    \textbf{a}-\textbf{d}, Device with $g_{1,\mathrm{off}}= \SI{-25}{\nano\metre}$ and $g_{2,\mathrm{off}}= \SI{0}{\nano\metre}$.
    The top row of plots shows experimentally measured transmission spectra ($T_{AA}(\lambda)$, $T_{BB}$, $T_{AB}$, and $T_{BA}$ respectively) in blue, while the second row shows the corresponding simulated spectra.
    The final row shows the extracted transmission and loss coefficients.
    \textbf{e}-\textbf{h}, Device with $g_{1,\mathrm{off}}= \SI{-25}{\nano\metre}$ and $g_{2,\mathrm{off}}= \SI{25}{\nano\metre}$. 
    The panels are organized in the same manner as in \textbf{a}-\textbf{d}.
    }
    \label{fig:SIfig4}
\end{figure*}

Spectral measurement is performed using a Keysight 81608A tunable laser source, which sweeps wavelengths from 1490 nm to 1640 nm with a fine 1 pm resolution to resolve narrow resonances. 
Light is edge-coupled into and out of the chip using lensed fibers with a $\SI{2.5}{\micro\metre}$ spot size, precisely aligned to the waveguide facets via high-resolution nano-positioning stages.
A manual fiber polarization controller ensures consistent input polarization for exciting fundamental TE mode.
For each measurement, the transmission spectra from both output ports are simultaneously recorded using a Keysight N7744A multi-channel optical power meter.

Figures~\ref{fig:SIfig3} and \ref{fig:SIfig4} present the transmission spectra for devices with engineered gap offsets, comparing experimental measurements (blue spectrum, top row) with simulations (black spectrum, second row).
A strong agreement is observed across all device configurations and transmission ports ($T_{AA}$, $T_{BB}$, $T_{AB}$, and $T_{BA}$). 
Key spectral features—including the resonant wavelengths, lineshapes, and extinction ratios—are consistently well-reproduced by our model, thereby validating its predictive accuracy. 
The bottom row of each figure panel shows the transmission and loss coefficients from simulations, which provide insight into the transmission zero condition and coupling regime transitions.

\subsection{Influence of the gap offset $g_{2,\mathrm{off}}$}

Figure~\ref{fig:SIfig3} isolates the impact of the secondary gap offset, $g_{2,\mathrm{off}}$, while the primary offset is kept constant at zero ($g_{1,\mathrm{off}}= \SI{0}{\nano\metre}$). 
A direct comparison between the device with $g_{2,\mathrm{off}}= \SI{50}{\nano\metre}$ (Fig.~\ref{fig:SIfig3}e-h) and the one with $g_{2,\mathrm{off}}= \SI{75}{\nano\metre}$ (Fig.~\ref{fig:SIfig3}a-d) reveals a consistent redshift in the resonant features across all spectra as $g_{2,\mathrm{off}}$ increases. 
This spectral shift is a direct consequence of the altered transmission coefficients ($t_{ij}$) in the three-waveguide coupler region. 
This behavior is reflected in the ratio of transmission coefficients (bottom row), which shows that the geometric change impacts the transmission coefficient terms, leading to the observed systematic shift.

\subsection{Influence of the gap offset $g_{1,\mathrm{off}}$}

Conversely, Figure~\ref{fig:SIfig4} demonstrates the role of gap offset $g_{1,\mathrm{off}}$ which directly controls the coupling strength between the two bus-waveguides (Waveguide $A$ and $B$). 
Introducing a negative offset of $g_{1,\mathrm{off}} = \SI{-25}{\nano\metre}$ (Fig.~\ref{fig:SIfig4}a-d) significantly reduces the gap between the interacting waveguides, thereby enhancing their evanescent coupling. 
This stronger interaction manifests as a much more dramatic, wavelength-dependent modulation of the spectral response.

This effect is particularly evident in the cross-port transmission spectra ($T_{AB}$ and $T_{BA}$), which exhibit sharp, high-extinction Fano resonances. 
This dynamic behavior is directly supported by the ratio of transmission coefficients (bottom row of Figs.~\ref{fig:SIfig4}), which show significantly larger variations in the ratio of transmission coefficients ($t_{21}/t_{12}$, $t_{12}/t_{21}$) compared to the devices shown in Fig.~\ref{fig:SIfig3}.

\newpage
\section{$Q$-factor Extraction}

\begin{figure*}[h!]
\centering
\includegraphics[width=1\textwidth]{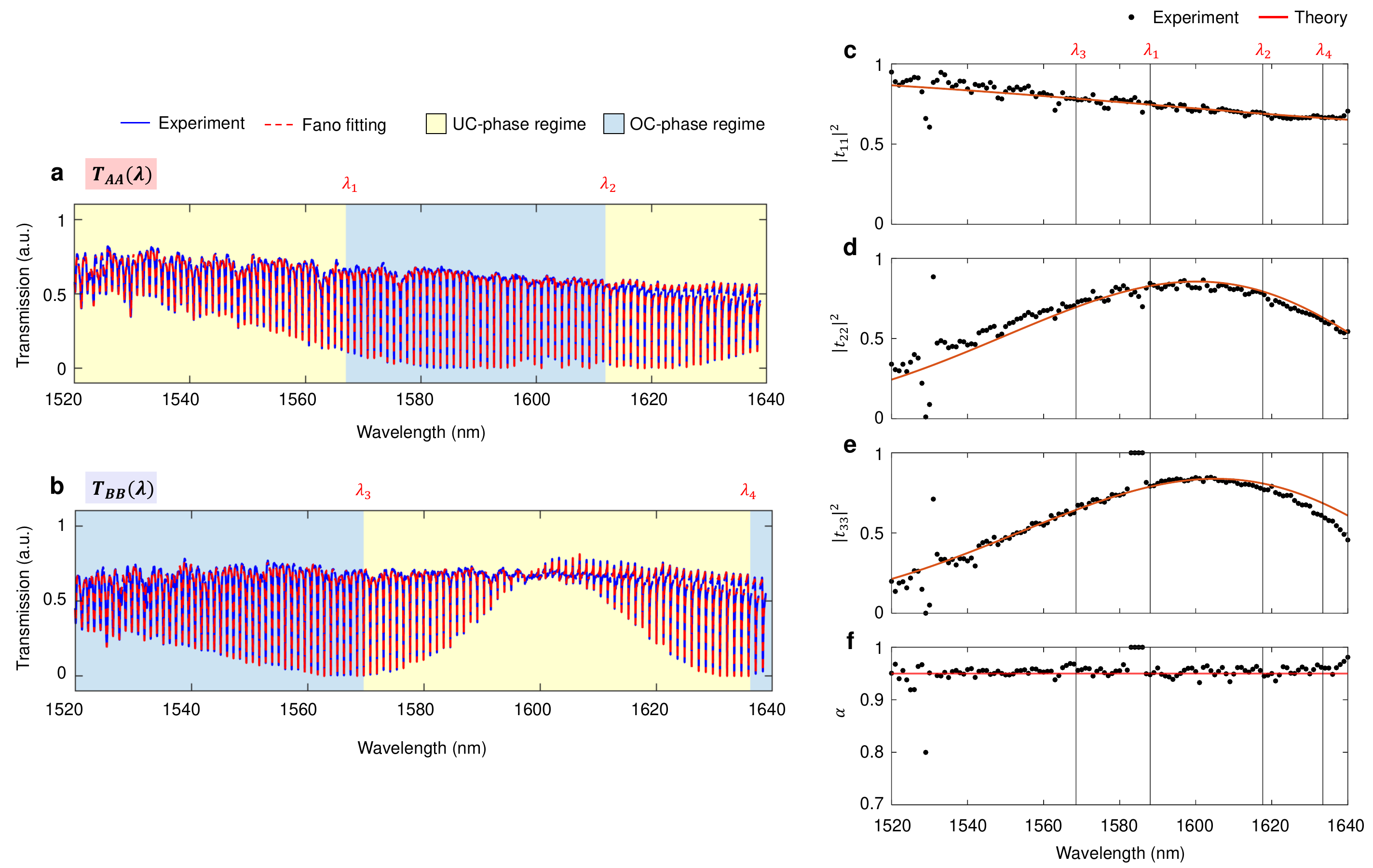}
\caption{
\textbf{Extraction of coupling and loss coefficients via Fano function fitting.}
\textbf{a}, \textbf{b}, Experimental transmission spectra for $T_{AA}(\lambda)$ (\textbf{a}) and $T_{BB}(\lambda)$ (\textbf{b}) are shown in blue. Each resonance is fitted using a general Fano function (red curves, Eq.~\ref{eq:Fanofunction}). Red markers ($\lambda_1$, $\lambda_2$, $\lambda_3$, $\lambda_4$) denote the wavelengths where transmission zero occurs. 
\textbf{c}, \textbf{d}, Wavelength-dependent transmission coefficients $|t_{11}|^2$ (\textbf{c}), $|t_{22}|^2$ (\textbf{d}), and $|t_{33}|^2$ (\textbf{e}) extracted from the fits (black points) compared with theoretical simulations (red curves).
\textbf{f}, Extracted round-trip loss coefficient (black points) as a function of wavelength, showing good agreement with the assumed constant value of 0.95 (red line).
}
\label{fig:SIfig5}
\end{figure*}
%%%%%%%%%%%%%%%%%%%%%%%%%%%%%%%%%%%%%%%%%%%%%%%%%%%%%%%%%%%%%%%%%%%%%%%%%%%%%%%%%%%%%%%%%%%%%

For a conventional single-bus ring resonator, the loaded quality factor ($Q_L$) can be extracted from the resonance linewidth, and the intrinsic ($Q_i$) and coupling quality factors ($Q_c$) can be estimated from the resonance transmission depth $T_{\mathrm{min}}$ using~\cite{Luo:11}:
\begin{equation}
Q_i = \frac{2Q_L}{1 \pm \sqrt{T_{\mathrm{min}}}}, \quad
Q_c = \frac{2Q_L}{1 \mp \sqrt{T_{\mathrm{min}}}},
\label{eq:ringQcalculation}
\end{equation}
where the sign is chosen according to the coupling regime: "$+$" for under-coupling, and "$-$" for over-coupling.

A similar approach applies to the dual-bus racetrack resonator. The $Q$-factors can be theoretically expressed as:
\begin{subequations}
\begin{align}
Q_L = \frac{\omega_0}{2\gamma_t} = \frac{\pi n_g L_{\mathrm{rt}}}{-\lambda_0 \ln|\alpha t_{33}|},
\label{eq:loadedQ}\\
Q_i = \frac{\omega_0}{2\gamma_i} = \frac{\pi n_g L_{\mathrm{rt}}}{-\lambda_0 \ln|\alpha|},
\label{eq:intrinsicQ}\\
Q_c = \frac{\omega_0}{2\gamma_c} = \frac{\pi n_g L_{\mathrm{rt}}}{-\lambda_0 \ln|t_{33}|},
\label{eq:couplingQ}
\end{align}
\end{subequations}
where $\gamma_t$, $\gamma_i$, and $\gamma_c$ are the total (loaded), intrinsic, and coupling amplitude decay rates, respectively.
$\lambda_0$ is the resonance wavelength, $n_g$ is the group index, and $L_{\mathrm{rt}}$ is the round-trip length of the resonator.

To extract $Q_L$ from the measured spectra, each resonance is fitted using a general Fano function \cite{PhysRevB.87.125118}:
\begin{equation}
T(\lambda) = p^2 \frac{\left( q + \dfrac{2(\lambda - \lambda_0)}{\gamma_t} \right)^2 + r^2}{1 + \dfrac{4(\lambda - \lambda_0)^2}{\gamma_t^2}},
\label{eq:Fanofunction}
\end{equation}
where $p^2$ is the non-resonant transmission intensity, $q$ is the classical Fano asymmetry factor, $r$ another lineshape factor that accounts for the ratio between Fano and Lorentzian, and $\gamma_t$ is the resonance linewidth. As shown in Figs.~\ref{fig:SIfig5}\textbf{a} and \textbf{b}, this function reliably fits both symmetric and asymmetric resonances in $T_{AA}(\lambda)$ and $T_{BB}(\lambda)$ across a broadband.

Extracting $\alpha$ to calculate $Q_i$ requires both $T_{AA}(\lambda)$ and $T_{BB}(\lambda)$. 
The parameters $t_{11}$, $t_{22}$, $t_{33}$, and $\alpha$ influence both transfer functions, and can be determined from four known quantities: the minimum transmission values $T_{AA,\mathrm{min}}$ and $T_{BB,\mathrm{min}}$, the non-resonant intensity $p^2$ of $T(\lambda)$, and the resonance linewidth $\gamma_t$. These quantities enable the evaluation of:
\begin{equation}
\pm \sqrt{T_{AA,\mathrm{min}}}
=
\frac{|t_{11}| - \alpha |t_{22}|}{1 - \alpha|t_{33}|},
\label{eq:signofTAA}
\end{equation}
\begin{equation}
\pm \sqrt{T_{BB,\mathrm{min}}}
=
\frac{|t_{22}| - \alpha |t_{11}|}{1 - \alpha|t_{33}|},
\label{eq:signofTBB}
\end{equation}
where the sign is chosen based on whether the resonance is in the UC or OC-phase regime. The sign convention for different wavelength ranges is summarized in Table~\ref{tab:signconvention}.

\begin{table}[htbp]
\centering
\begin{tabular}{@{}c@{\hskip 20pt}c@{\hskip 20pt}c@{}}
\toprule
\textbf{Wavelength range} & sign of \(\pm \sqrt{T_{AA,\mathrm{min}}(\lambda)}\) & sign of \(\pm \sqrt{T_{BB,\mathrm{min}}(\lambda)}\) \\
\midrule
\(\lambda < \lambda_3\)                & \textbf{$+$} & \textbf{$-$} \\[3ex]
\(\lambda_3 < \lambda < \lambda_1\)    & \textbf{$+$} & \textbf{$+$} \\[3ex]
\(\lambda_1 < \lambda < \lambda_2\)    & \textbf{$-$} & \textbf{$+$} \\[3ex]
\(\lambda_2 < \lambda < \lambda_4\)    & \textbf{$+$} & \textbf{$+$} \\[3ex]
\(\lambda > \lambda_4\)                & \textbf{$+$} & \textbf{$-$} \\
\bottomrule
\end{tabular}
\caption{Sign conventions used to determine $\pm \sqrt{T_{AA,\mathrm{min}}}$ and $\pm \sqrt{T_{BB,\mathrm{min}}}$ in each wavelength regime.}
\label{tab:signconvention}
\end{table}

The transmission coefficient $|t_{33}(\lambda)|$ is obtained from Eq.~\ref{eq:loadedQ}, while $|t_{11}(\lambda)|$ is given by the fitted $p^2$ from $T_{AA}(\lambda)$. With the sign conventions specified, Eqns.~\ref{eq:signofTAA} and \ref{eq:signofTBB} form a system of linear equations that can be solved for $|t_{22}(\lambda)|$ and $\alpha(\lambda)$.
The results of this extraction, and their comparison with theoretical simulations, are shown in Figs.~\ref{fig:SIfig5}\textbf{c}–\textbf{f}.

\textit{Note:} While $|t_{22}(\lambda)|$ can alternatively be obtained by fitting $T_{BB}(\lambda)$ using the Fano function, this approach introduces greater baseline-related error. 
Hence, we adopt the method based on the coupled equations for improved accuracy.

\newpage
\section{Experimental Results for \SiN~Device}

 \begin{figure}[htbp]
 	\centering
 	\includegraphics[width=0.5\columnwidth]{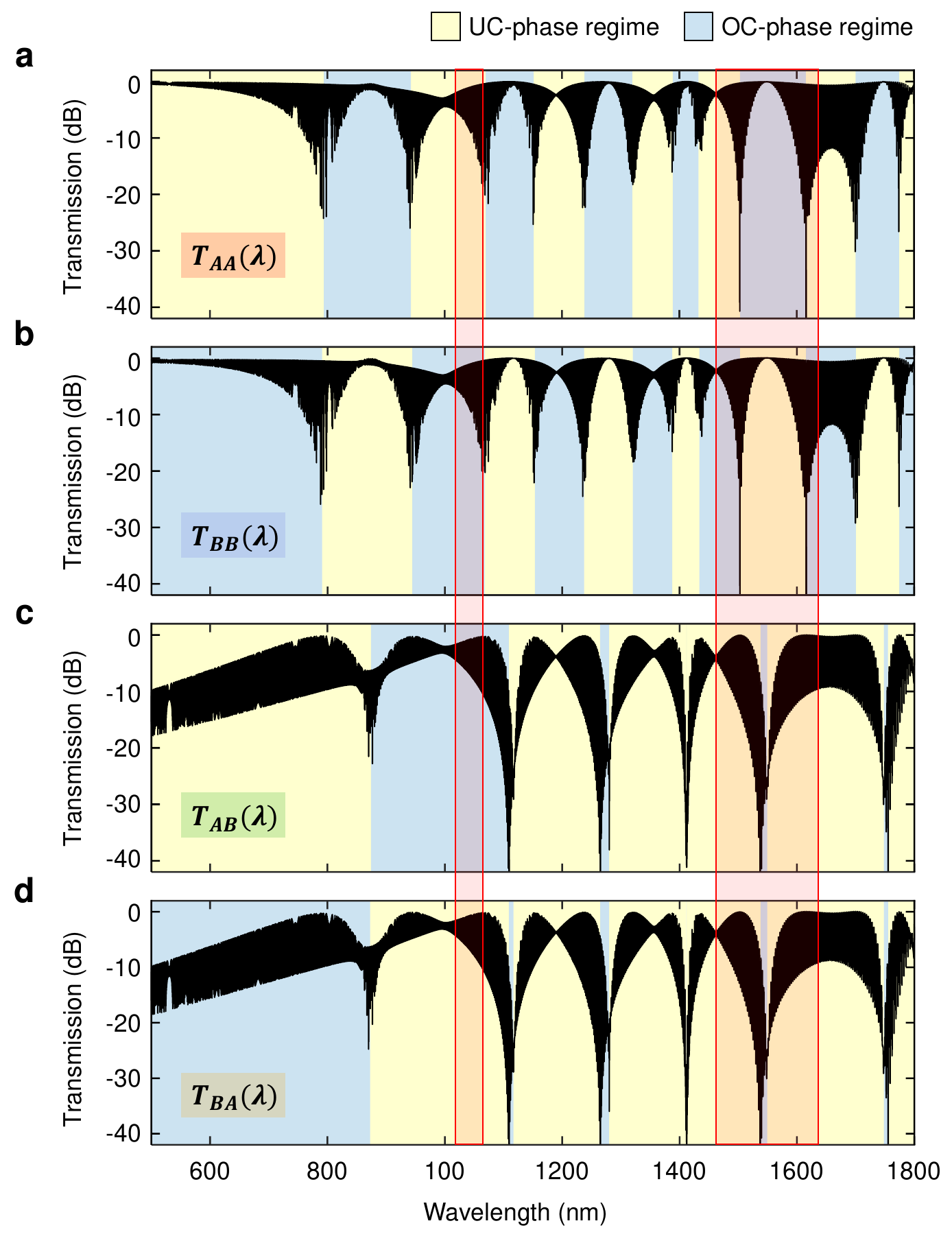}
 	\caption{
 	\textbf{Simulated transmission spectra of the dual-bus racetrack resonator on the \SiN~platform over a broad wavelength range (500–1800~nm).}
 	   \textbf{a}, Transmission spectrum $T_{AA}(\lambda)$. 
 	   \textbf{b}, $T_{BB}(\lambda)$. 
 	   \textbf{c}, $T_{AB}(\lambda)$. 
 	   \textbf{d}, $T_{BA}(\lambda)$.
 	The red boxes indicate the experimentally accessible wavelength ranges (1010–1090~nm and 1450–1650~nm), limited by the available tunable laser sources.
 	}
 	\label{fig:SIfig6}
     \vspace{1mm}
 \end{figure}
%%%%%%%%%%%%%%%%%%%%%%%%%%%%%%%%%%%%

\begin{figure}[htbp]
    \centering
    \includegraphics[width=1\textwidth]{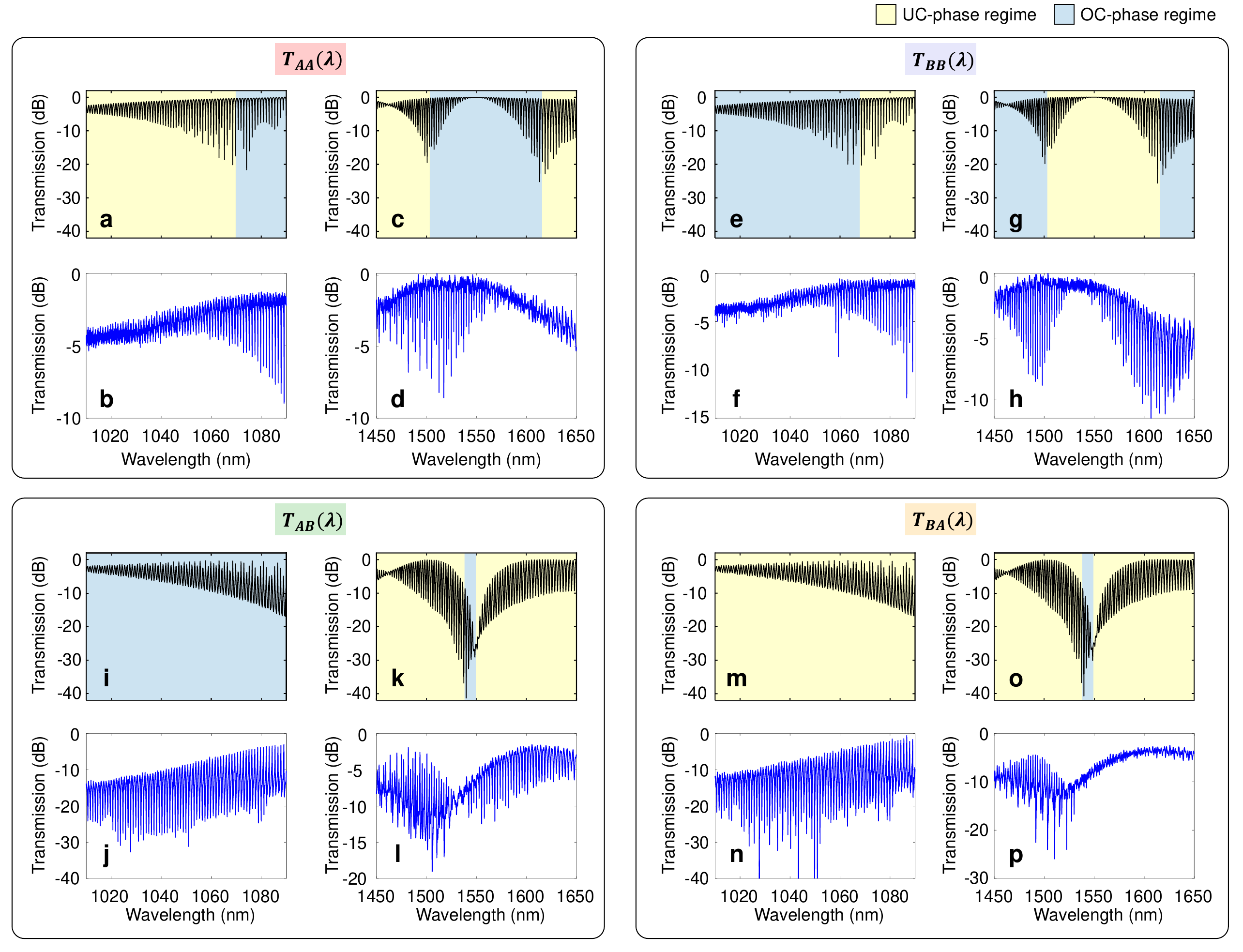}
    \caption{
    \textbf{Simulated and experimental transmission spectra of the \SiN~dual-bus resonator over two wavelength bands (1010–1090 nm and 1450–1650 nm).}
    \textbf{a–d}, Simulated (\textbf{a}, \textbf{c}) and experimental (\textbf{b}, \textbf{d}) transmission spectra $T_{AA}(\lambda)$ in the wavelength ranges 1010–1090 nm (\textbf{a}, \textbf{b}) and 1450–1650 nm (\textbf{c}, \textbf{d}).
    \textbf{e–h}, Simulated (\textbf{e}, \textbf{g}) and experimental (\textbf{f}, \textbf{h}) transmission spectra $T_{BB}(\lambda)$ in the same two bands.
    \textbf{i–l}, Simulated (\textbf{i}, \textbf{k}) and experimental (\textbf{j}, \textbf{l}) transmission spectra $T_{AB}(\lambda)$.
    \textbf{m–p}, Simulated (\textbf{m}, \textbf{o}) and experimental (\textbf{n}, \textbf{p}) transmission spectra $T_{BA}(\lambda)$.
    In all plots, black curves correspond to simulations and blue curves to measurements.
    }
    \label{fig:SIfig7}
\end{figure}

Broadband simulations (500-1800 nm) of the dual-bus racetrack resonator on the \SiN~platform confirm its characteristic coupling features, including the transition between complementary UC and OC-phase regimes and the presence of Fano resonance lineshapes (Fig.~\ref{fig:SIfig6}).
The dual-bus racetrack resonator was fabricated on a \SiN~platform. 
The device used a 300~nm-thick \SiN~ layer and a waveguide width of 700~nm, ensuring single-mode ($\text{TE}_0$) operation. 
All geometrical parameters—including coupling gaps and resonator radius—is the same as the silicon-on-insulator (SOI) device.

The results in Fig.~\ref{fig:SIfig7} correspond to a device with gap offsets $g_{1,\mathrm{off}}= \SI{0}{\nano\metre}$ and $g_{2,\mathrm{off}}= \SI{100}{\nano\metre}$. 
Simulated (black) and experimental (blue) transmission spectra are shown across two wavelength ranges: 1010–1090~nm and 1450–1650~nm. Overall, the measured spectra exhibit trends consistent with simulation, although some discrepancies are observed.
These discrepancies likely arise from fabrication-induced imperfections and limitations in simulation assumptions.
Particularly the uniform round-trip loss coefficient $\alpha$, which was set to 0.99 across all wavelengths. 
In reality, the propagation loss of \SiN~ varies significantly with wavelength, especially due to hydrogen-related absorption peaks inherent to the PECVD process~\cite{Liu:24}.

Despite these differences, the measured spectra clearly capture key features predicted by theory.
In particular, extinction ratio variation across the spectrum reflects the expected coupling regime changes. 
Lorentzian resonance lineshapes are observed in the direct-port transmissions $T_{AA}(\lambda)$ and $T_{BB}(\lambda)$, while cross-port transmissions $T_{AB}(\lambda)$ and $T_{BA}(\lambda)$ exhibit wavelength-dependent transitions from Lorentzian dips to peaks and strong Fano asymmetries.

\bibliographystyle{naturemag}

\end{document}